\documentclass[aps,prc,twocolumn,groupedaddress,nofootinbib]{revtex4-2}

\usepackage{amsmath,amssymb,mathtools}
\usepackage{xcolor}
\usepackage{booktabs}
\usepackage{graphicx}
\usepackage{siunitx}
\usepackage{url}
\usepackage{hyperref}
\hypersetup{colorlinks=true, allcolors=blue}
\usepackage{orcidlink}
\usepackage{tikz}
\usepackage{pgfplots}
\pgfplotsset{compat=1.18}
\usepackage[switch]{lineno}
\usepackage{subcaption}
\usepackage[stable]{footmisc}
\usepackage[font=small,labelfont=bf]{caption}
\usepackage{caption}
\usepackage{textcomp}
\definecolor{darkgreen}{RGB}{0,128,0}

\begin{document}
%%\linenumbers
\modulolinenumbers[5]
\setlength{\linenumbersep}{4pt}

\title{Measurements of Bremsstrahlung-Averaged Cross Sections for Reactions on Natural Nickel Targets at $E_{\mathrm{endpoint}}=\SI{40}{\mega\electronvolt}$}
\author{O.~Nusair\,\orcidlink{0000-0002-4841-5286}}
\email{onusair@northstarnm.com}
\thanks{Corresponding author}

\author{N.~Solomon}

\author{M.~Toro-Gonzalez\,\orcidlink{0000-0003-3320-423X}}

\author{D.~DeVries}

\affiliation{NorthStar Medical Radioisotopes, LLC, Beloit, Wisconsin 53511, USA}

\date{\today}

\begin{abstract}
Bremsstrahlung activation measurements were performed to study the production of
$^{57}\mathrm{Ni}$, $^{56}\mathrm{Ni}$,  $^{58}\mathrm{Co}$, $^{57}\mathrm{Co}$,
$^{56}\mathrm{Co}$, and $^{55}\mathrm{Co}$ from natural nickel targets irradiated
with photons generated by \SI{40}{MeV} electrons incident on a tantalum converter.
Bremsstrahlung spectra were modeled using MCNP6.3\texttrademark{} and experimentally validated
through activation of natural tin. Bremsstrahlung-averaged cross sections were
extracted from end-of-irradiation activities, yielding
$\langle\sigma\rangle = 8.983~\pm~0.028$~mb for
$^{58}\mathrm{Ni}(\gamma,n)^{57}\mathrm{Ni}$,
$0.248~\pm~0.025$~mb for
$^{58}\mathrm{Ni}(\gamma,2n)^{56}\mathrm{Ni}$,
$0.704~\pm~0.218$~mb for
$^{nat}\mathrm{Ni}(\gamma,pxn)^{58}\mathrm{Co}$, $9.192~\pm~0.386$~mb for
$^{58}\mathrm{Ni}(\gamma,p)^{57}\mathrm{Co}$, and
$2.239~\pm~0.355$~mb for
$^{58}\mathrm{Ni}(\gamma,pn)^{56}\mathrm{Co}$.
A $90\%$ confidence-level upper limit of
$\langle\sigma\rangle < 0.021$~mb is established for the
$^{58}\mathrm{Ni}(\gamma,p2n)^{55}\mathrm{Co}$ channel.
Comparison with JENDL-5 evaluations and prior studies indicates channel-dependent agreement, 
with residual discrepancies observed for selected charged-particle emission reactions. 
In particular, the measured $^{58}\mathrm{Ni}(\gamma,p)^{57}\mathrm{Co}$ cross section exceeds the JENDL-5 prediction by approximately a factor of two, whereas TALYS-2.2 calculations reproduce the experimental value, suggesting an underestimation of the $(\gamma,p)$ channel strength in JENDL-5 under bremsstrahlung conditions. 
For the $^{58}\mathrm{Ni}(\gamma,pn)^{56}\mathrm{Co}$ channel, both TALYS-2.2 and JENDL-5 predictions are in reasonable agreement, while the present measurement is higher by approximately a factor of two, though with larger associated uncertainty.
\end{abstract}

\maketitle

\section{Introduction}
\label{sec:intro}

Photonuclear activation experiments employing bremsstrahlung photon fields provide access to reaction yields 
integrated over a continuous photon energy distribution extending from the reaction threshold up to the endpoint 
energy defined by the incident electron beam. Such measurements are relevant both for medical radioisotope 
production and for benchmarking photonuclear reaction modeling and evaluated nuclear data libraries, including 
modern general-purpose evaluations and dedicated photonuclear databases~\cite{Brown2018,Koning2019,Kawano2020,Iwamoto2023}.

Activation of nickel and cobalt targets leads to the production of radionuclides with well-established 
roles in nuclear medicine. In particular, $^{55}\mathrm{Co}$ ($T_{1/2}\approx17.53$~h) is a positron emitter 
that has been investigated for positron emission tomography (PET) applications, while both $^{57}\mathrm{Co}$ 
($T_{1/2}\approx271.8$~d) and $^{56}\mathrm{Co}$ ($T_{1/2}=77.236$~d) are widely used for $\gamma$-ray spectrometry 
calibration, with $^{57}\mathrm{Co}$ serving as a standard reference source~\cite{Spellerberg1998} and $^{56}\mathrm{Co}$ 
providing high-energy efficiency calibration due to its rich and intense $\gamma$-ray cascade~\cite{Miyahara1998}. These 
applications motivate continued efforts to quantify production yields and bremsstrahlung-averaged cross sections under 
production-relevant irradiation conditions and to improve the experimental database for nickel-based targets~\cite{Khandaker2011,Deiev2022NiPhotonuclear}.

Photon-induced production of $^{58}\mathrm{Co}$ ($T_{1/2}=70.883$~d) on nickel proceeds through 
charged-particle emission channels that open above the giant dipole resonance (GDR) and are sensitive to the 
balance between direct and compound-nucleus mechanisms. Experimental information on these channels remains limited, 
and existing bremsstrahlung-based measurements show notable deviations from model predictions, particularly those 
obtained with default TALYS~\cite{Koning2023} inputs~\cite{Zaman2018NatNiPhotonuclear,Naik2020NatNiCo58,Deiev2022NiPhotonuclear,Timchenko2025Co58Photonuclear}. 
Additional cross-section data are therefore required to better constrain reaction modeling and improve evaluated photonuclear databases.

Reliable extraction of such yields and flux-averaged observables places stringent requirements on the 
characterization of the incident photon field. This requirement is particularly critical in bremsstrahlung 
activation experiments, where measured quantities represent integrals over a broad energy spectrum and are 
frequently compared to evaluated data. Consequently, validation of the bremsstrahlung photon-flux model 
constitutes a necessary prerequisite for quantitative interpretation of nickel activation results.

Fortunately, extensive experimental photonuclear datasets exist for elements with multiple naturally abundant 
isotopes, including elemental tin~\cite{Varlamov2010SnPhotonuclear,Fultz1969,Utsunomiya2011}. Tin therefore 
provides a robust basis for validating bremsstrahlung flux models. In the present work, tin data are leveraged to 
validate the Monte Carlo (MC) photon-flux model prior to its application to activation measurements on natural nickel.

For a target, such as natural nickel, containing $N_t$ atoms of the relevant isotope and irradiated with a 
constant bremsstrahlung photon flux, the production rate $R$ of a radioactive product nucleus is given by
\begin{equation}
R =
N_t
\int_{E_{\mathrm{thr}}}^{E_0}
\sigma(E_\gamma)\,
\Phi(E_\gamma)\,
dE_\gamma ,
\label{eq:production_rate}
\end{equation}
where $\sigma(E_\gamma)$ is the photonuclear cross section, $\Phi(E_\gamma)$ is the differential photon 
flux, $E_{\mathrm{thr}}$ is the reaction threshold energy, and $E_0$ is the incident electron beam energy, also known as the bremsstrahlung end-point energy.

During irradiation, radioactive nuclei decay concurrently with their production according to their half-lives. 
Assuming a constant production rate $R$, the activity at the end of irradiation (EOI) is
\begin{equation}
A_{\mathrm{EOI}} =
R \left( 1 - e^{-\lambda t_{\mathrm{irr}}} \right),
\label{eq:activity_eoi}
\end{equation}
where $\lambda = \ln 2 / T_{1/2}$ is the decay constant associated with the half-life $T_{1/2}$ of the product 
nucleus and $t_{\mathrm{irr}}$ is the irradiation time.

Substituting Eq.~(\ref{eq:production_rate}) into Eq.~(\ref{eq:activity_eoi}) yields
\begin{equation}
A_{\mathrm{EOI}} =
N_t
\left( 1 - e^{-\lambda t_{\mathrm{irr}}} \right)
\int_{E_{\mathrm{thr}}}^{E_0}
\sigma(E_\gamma)\,
\Phi(E_\gamma)\,
dE_\gamma .
\label{eq:activity_full}
\end{equation}

Moreover, the bremsstrahlung-averaged cross section (BACS) is defined as the photon-flux–weighted average of 
the energy-dependent reaction cross section, integrated over the bremsstrahlung energy distribution from the 
reaction threshold to the endpoint energy, and expressed as:
\begin{equation}
\langle \sigma \rangle_{\mathrm{BACS}}
=
\frac{\displaystyle
\int_{E_{\mathrm{thr}}}^{E_0}
\sigma(E_\gamma)\, \Phi(E_\gamma)\, dE_\gamma}
{\displaystyle
\int_{E_{\mathrm{thr}}}^{E_0}
\Phi(E_\gamma)\, dE_\gamma } .
\label{eq:bacs_fraction}
\end{equation}
Using this definition, the BACS can be extracted from measured EOI activities according to
\begin{equation}
\langle \sigma \rangle_{\mathrm{BACS}}
=
\frac{A_{\mathrm{EOI}}}
{N_t
\left( 1 - e^{-\lambda t_{\mathrm{irr}}} \right)
\displaystyle
\int_{E_{\mathrm{thr}}}^{E_0}
\Phi(E_\gamma)\, dE_\gamma } .
\label{eq:bacs_from_activity}
\end{equation}

Based on Eq.~(\ref{eq:bacs_from_activity}), determination of bremsstrahlung-averaged cross sections 
requires two key inputs: (i) an accurate evaluation of the bremsstrahlung photon flux integrated between 
the reaction threshold and endpoint energy, and (ii) a precise determination of the EOI activity of the reaction product.

This paper is organized in two parts with the objective of determining bremsstrahlung-averaged cross 
sections for several photonuclear reactions on a natural nickel target. In the first part, bremsstrahlung photon 
fluxes are evaluated using MCNP6.3~\cite{MCNP63} simulations for a realistic experimental geometry implemented 
through the Attila4MC\texttrademark{} interface~\cite{SilverFirAttila2024}. An experimental validation of the 
simulated flux is then performed using a $\chi^2$-minimization procedure applied to a parameterized two-energy-group 
photon flux, constrained to reproduce EOI activities measured from the activation of a natural tin target.

In the second part, the measured EOI activities of six reaction products, together with the validated 
integrated bremsstrahlung photon flux, are used to extract BACS for the photonuclear reactions
$^{58}\mathrm{Ni}(\gamma,n)^{57}\mathrm{Ni}$,
$^{58}\mathrm{Ni}(\gamma,2n)^{56}\mathrm{Ni}$,
$^{nat}\mathrm{Ni}(\gamma,pxn)^{58}\mathrm{Co}$,
$^{58}\mathrm{Ni}(\gamma,p)^{57}\mathrm{Co}$,
$^{58}\mathrm{Ni}(\gamma,pn)^{56}\mathrm{Co}$, and
$^{58}\mathrm{Ni}(\gamma,p2n)^{55}\mathrm{Co}$.

\section{Experimental Method, Bremsstrahlung Spectrum Modeling and Experimental Validation of the Model}
\label{sec:converter}

The main nickel target irradiation was performed using the TT-300HE Rhodotron\texttrademark{} electron accelerator at
NorthStar Medical Radioisotopes, LLC (NMR), in Beloit, WI. The electron beam was operated at an
energy of \SI{40}{\mega\electronvolt} with a nominal average beam current of \SI{1}{\micro\ampere}, noting limited precision in the beam current determination at such low values. Bremsstrahlung photons were produced by directing the
electron beam onto a water-cooled tantalum converter of
\SI{6}{\milli\meter} thickness.
For \SI{40}{\mega\electronvolt} electrons, the continuous
slowing-down approximation (CSDA) range in tantalum is approximately
\SI{8.2}{\milli\meter} \cite{NISTSTAR2005}, corresponding to a relative thickness
$t/R_{\mathrm{CSDA}} \approx 0.73$. At this thickness, electrons undergo
substantial energy loss and multiple bremsstrahlung emission events, resulting
in spectral softening relative to an ideal thin-target spectrum.

The activation target consisted of a high-purity natural nickel foil
(purity \SI{99.99}{\percent}) with a mass of \SI{0.44919}{g}.
The foil had a thickness of \SI{0.5}{\milli\meter} and a diameter of
\SI{13}{\milli\meter}. During irradiation, the nickel foil was mounted directly
against the thin aluminum back window of the converter assembly to ensure
reproducible geometry and maximal photon fluence at the target position.
The irradiation duration was \SI{61}{s}.

Following the end of irradiation, the activated nickel foil was transferred to a low-background counting 
station equipped with a high-purity germanium (HPGe) detector. Gamma-ray counting commenced
\SI{1.37}{h} after the end of irradiation and continued for approximately \SI{14}{h} of real time, with a measured fractional dead time of
\SI{0.22}{\percent}.

The determination of the EOI activities of the reaction products,
which form the basis for the extraction of bremsstrahlung-averaged cross sections,
is discussed in Sec.~\ref{sec:cross sections}. However, a prerequisite for this analysis
is an accurate and validated description of the bremsstrahlung photon flux incident on
the activation target. We therefore first describe the MC modeling of the
photon flux and the associated validation methodology, which underpins the subsequent
activity and cross-section determinations.

\subsection{Monte Carlo simulation and photon-flux representation}
\label{subsec:mc_framework}
\begin{figure}[t]
    \centering
    \includegraphics[width=0.85\columnwidth]{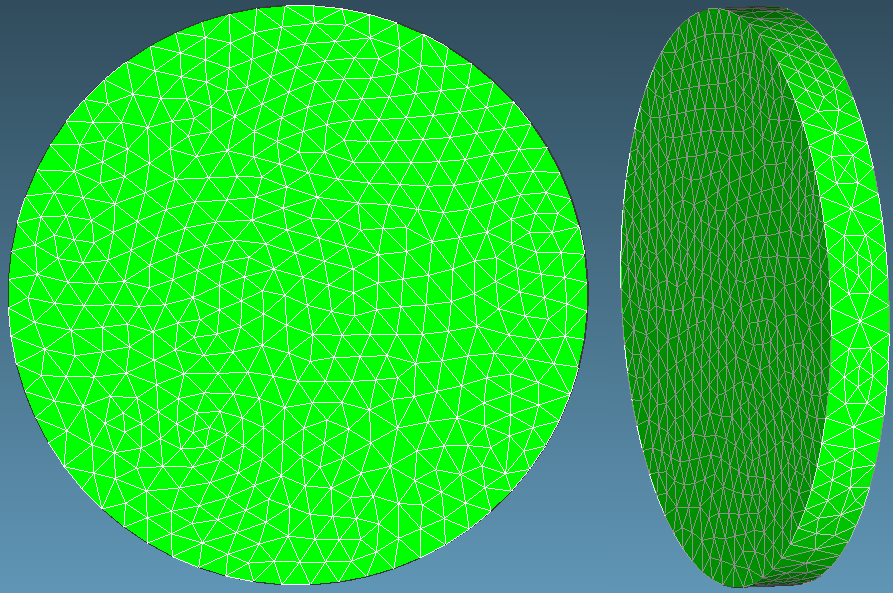}
    \caption{Meshed tin target geometry generated using Attila4MC. The front view illustrates the overall 
	cylindrical geometry and surface discretization, while the tilted side view highlights the target 
	thickness and tetrahedral mesh refinement through the volume.}
    \label{fig:meshed_tin_target}
\end{figure}

The bremsstrahlung photon flux incident on the activation targets was modeled
using the Monte Carlo radiation transport code MCNP6.3. The full converter, target (illustrated in 
Fig.~\ref{fig:meshed_tin_target}), and shielding geometry were represented explicitly using an 
unstructured three-dimensional mesh generated from a detailed CAD model and imported into MCNP6.3 via the Attila4MC embedded-geometry interface.

The computational model comprises 132 distinct geometric regions corresponding to
individual mechanical components of the irradiation assembly. These include the
tantalum bremsstrahlung converter, aluminum structural elements, water-cooling
channels, stainless-steel shielding components, and the activation target.

For validation of the simulated photon flux, we modeled a natural tin target 
using the same experimental geometry and beam conditions as those employed in the nickel
irradiation. In this validation configuration, a \SI{10}{s} irradiation was
simulated with a \SI{40}{MeV} electron beam at a current of \SI{1}{\micro A}
incident on a \SI{0.6186}{g} high-purity natural tin target. The tin composition
was taken as 99.3\% Sn with a 0.7\% Cu impurity fraction. Additional details
regarding the experimental tin activation measurements are provided in ~\cite{Nusair2026}.

The natural tin target was modelled as a dedicated region corresponding to a
right circular cylinder with a nominal radius of \SI{5}{mm} and a height of
\SI{1.0784}{mm}. The CAD-derived target volume
($0.0846972~\mathrm{cm^{3}}$) was reproduced by the tetrahedral mesh to within
$1.05\times10^{-4}~\mathrm{cm^{3}}$, with a final meshed volume of
$0.0845924~\mathrm{cm^{3}}$. The target region was discretized into 6{,}907
tetrahedral cells, indicating good agreement between the CAD geometry and its
numerical representation.

The unstructured computational mesh of the tin flux monitor contains approximately $9.9\times10^{5}$
tetrahedral elements and was generated using the Attila4MC interface.
All geometric components were imported directly from CAD models and discretized
with the Attila4MC mesher using default settings, with global edge-length controls
of \SIrange{0.0125}{0.05}{cm}. Curvature-based mesh refinement was enabled using a tolerance of
\(d/h = 0.020\), where \(d\) is the maximum deviation between the CAD
surface and the mesh representation and \(h\) is the local mesh edge
length, ensuring adequate geometric resolution in regions of high
curvature.

Material definitions employed isotopic compositions and mass densities consistent
with engineering specifications. The tin target used a natural isotopic
composition, while the tantalum bremsstrahlung converter was modelled as elemental
Ta with a density of $16.68~\mathrm{g\,cm^{-3}}$. Water regions were assigned
thermal scattering treatment using ENDF/B-VIII.0 S($\alpha,\beta$) data~\cite{Brown2018} for
hydrogen in light water at 300~K.

\subsection{Particle transport physics}

Coupled electron--photon--neutron transport was enabled using
\texttt{mode n p e}. Electron transport included bremsstrahlung production and
energy loss processes relevant to high-energy electron beams, with electron
physics parameters selected to ensure accurate modelling of bremsstrahlung
generation in the thick tantalum converter. Photonuclear reactions were enabled
where applicable, while photonuclear interactions on hydrogen were explicitly
disabled. Neutron and photon physics options were chosen to ensure consistency
with evaluated nuclear data libraries used elsewhere in this work.

All geometric regions were assigned unit particle importance for electrons,
photons, and neutrons, with a surrounding vacuum region assigned zero importance
to terminate particle histories exiting the problem domain.

\subsection{Electron source definition}

The incident electron beam was modeled as a forward-directed monoenergetic source with an energy of 
\SI{40}{MeV}. The source was defined with a fixed beam
direction perpendicular to the converter and with spatial distributions in
the transverse directions chosen to reproduce the beam footprint at the
converter entrance. The source position distributions in the transverse directions were implemented 
using uniform sampling over the active beam area.
A total of $2\times10^{6}$ primary electron histories were simulated for each
calculation. Three independent simulations were performed corresponding to
Gaussian beam spot diameters (FWHM) of \SI{0.5}{\milli\meter},
\SI{4}{\milli\meter}, and \SI{6}{\milli\meter}, as discussed in Sec.~\ref{sec:mc_model}. The goal of 
these simulations is to quantify the impact of beam spot diameter on the final photon flux. 

\subsection{Photon flux tallies}

The bremsstrahlung photon flux was tallied within the tin target region using an
MCNP6.3 cell-averaged photon flux tally (\texttt{F4:p}). The tally was defined over
an energy grid spanning \SIrange{1}{40}{MeV}, with 38 linearly spaced
energy bins below the endpoint energy and a final upper bin extending to the
\SI{40}{MeV} endpoint. The resulting tally output represents per-bin photon flux
integrals per incident electron, which are directly compatible with subsequent
activation and bremsstrahlung-averaged cross-section calculations.

Tallies were written periodically during the simulation to monitor statistical
convergence, and the final results were extracted from the MCNP6.3 \texttt{MCTAL}
output file for post-processing and rebinning.

\subsection{Statistical considerations and simulation results}
\label{sec:mc_model}

The number of simulated histories was selected to ensure that the relative
statistical uncertainty of the photon flux in the energy region relevant for
photonuclear activation was below the few-percent level. Residual statistical
uncertainties from the MC calculation were propagated through the
subsequent flux-scaling and cross-section extraction procedures, as described in
Sec.~\ref{subsec:uncertainty}.

The bremsstrahlung photon flux was tallied and exported as energy-binned data.
The MC photon flux output is interpreted strictly as a set of
normalized per-bin integrals per incident electron, rather than as normalized
differential energy densities.

\begin{figure}[t]
  \centering
  \includegraphics[width=\linewidth]{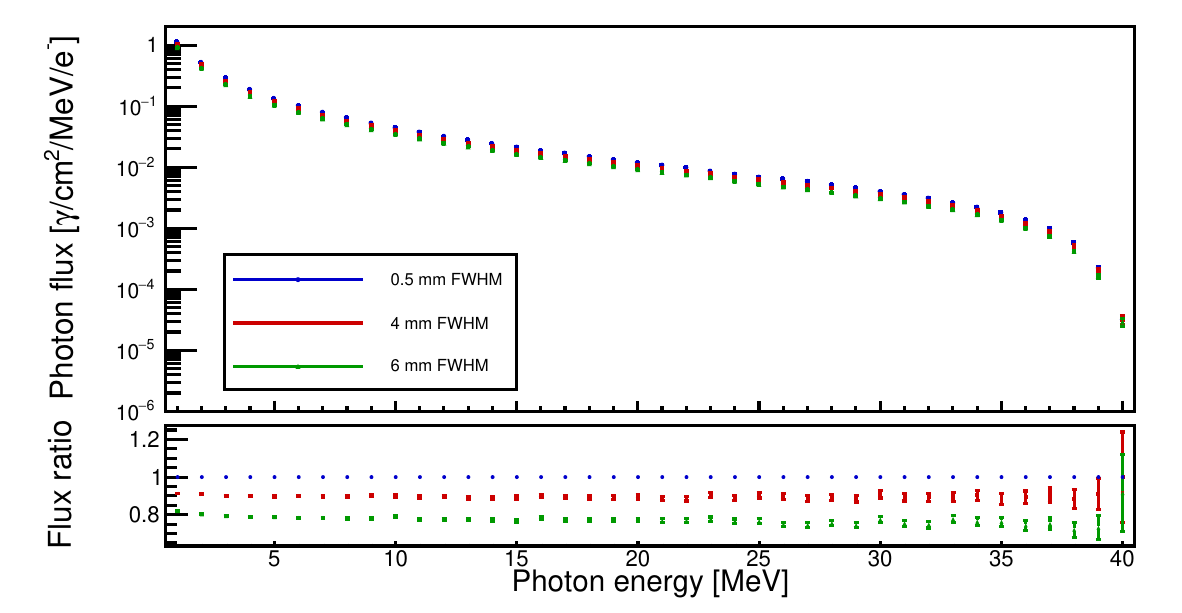}
  \caption{MCNP6.3 simulated bremsstrahlung photon fluxes at the tin target location as a function of 
  photon energy for beam spot diameters (FWHM) of \SI{0.5}{\milli\meter}, \SI{4}{\milli\meter}, and 
  \SI{6}{\milli\meter}. The fluxes are shown as per-bin integrals normalized per incident electron. 
  The bottom pad shows the ratios with respect to the \SI{0.5}{\milli\meter} flux.}
  \label{fig:flux}
\end{figure}

Figure~\ref{fig:flux} compares the bremsstrahlung photon flux at the tin target (flux monitor)
location as a function of photon energy for beam spot diameters (FWHM) of
\SI{0.5}{\milli\meter}, \SI{4}{\milli\meter}, and \SI{6}{\milli\meter}. This comparison is made because 
the beam shape and spot size can affect the activation outcome. However, no statistically significant
difference is observed between the three cases over the full energy range,
indicating that the simulated photon flux at the target is insensitive to the
beam spot size within this range.

\subsection{MC Flux Validation Using Tin Activation}

This section describes the physics model and numerical procedure used to validate
the MC bremsstrahlung photon flux through activation measurements
performed with a natural tin target. The validation is formulated in terms of two
multiplicative correction factors, or calibration parameters, that rescale the MC photon flux in two energy
intervals, \SIrange{6}{15}{MeV} and \SIrange{15}{40}{MeV}. These factors, denoted
$p_{\mathrm{lo}}$ and $p_{\mathrm{hi}}$, are determined by minimizing a $\chi^{2}$
constructed from measured end-of-irradiation activities of tin activation
products and the corresponding forward-model predictions obtained using the
corrected MC flux.

The natural tin target activation provided at least five independent measurements of initial
activities, enabling a constrained validation of the photon flux. For this
purpose, the MC flux is partitioned into three energy regions. The lowest-energy
region, corresponding to photon energies below the common threshold for single
nucleon removal (taken here as \SI{6}{MeV}), does not contribute to the observed
activation products and is therefore excluded from the $\chi^{2}$ minimization.
The remaining two regions are assigned independent normalization factors that
rescale the MC flux, $\Phi^{\mathrm{MC}}$, according to
\begin{equation}
\label{eq:corrected_flux}
\Phi^{\mathrm{corr}}(E)=
\begin{cases}
p_{\mathrm{lo}}\,\Phi^{\mathrm{MC}}(E), & E \in \mathcal{B}_{\mathrm{lo}}, \\[3pt]
p_{\mathrm{hi}}\,\Phi^{\mathrm{MC}}(E), & E \in \mathcal{B}_{\mathrm{hi}}, \\[3pt]
\Phi^{\mathrm{MC}}(E), & \text{otherwise},
\end{cases}
\end{equation}
with the low-energy band $\mathcal{B}_{\mathrm{lo}} = \SIrange{6}{15}{MeV}$ and the high-energy band
$\mathcal{B}_{\mathrm{hi}} = \SIrange{15}{40}{MeV}$.

All subsequent activity calculations and integrals are performed using the
corrected flux $\Phi^{\mathrm{corr}}$. Outside the energy interval
\SIrange{6}{40}{MeV}, the MC photon flux remains unscaled.

The five activation products measured in the activated natural tin and used in the $\chi^2$ minimization are 
listed in Table~(\ref{tab:fit_observables}) with their production routes and measured EOI activities. Details 
on individual peak fitting and the time-series analysis performed to extract their activities can be found in~\cite{Nusair2026}.

\begin{table}[h!]
 \caption{Observables used in the $\chi^2$ minimization method.}
  \centering
  \sisetup{round-mode=places,round-precision=3}
  \begin{tabular}{l l l  l}
    \hline
    Reaction  & $T_{1/2}$ & $A_{EOI}$ [Bq] & $\pm \sigma_{A_{EOI}}$ [Bq] \\
    \hline
    $^{112}$Sn($\gamma,2n$)$^{110}$Sn    & \SI{4.154}{h}    & \num{1735.580} & \num{11.4128}  \\
    $^{112}$Sn($\gamma,n$)$^{111}$Sn     & 35.3 m     & \num{1.988e5}  & \num{6.460e3}  \\
    $^{nat}$Sn($\gamma,xn$)$^{113}$Sn    & 115.08 d   & \num{34.520}   & \num{0.09556} \\
    $^{nat}$Sn($\gamma,xn$)$^{117m}$Sn   & 13.94 d    & \num{1007.270} & \num{0.899367} \\
    $^{124}$Sn($\gamma,n$)$^{123m}$Sn    & 40.06 m    & \num{9.428e5}  & \num{2.798e3}   \\    
    \hline
  \end{tabular}
  \label{tab:fit_observables}
\end{table}
For $^{111}$Sn, the initial activity was extracted using four independent
$\gamma$-ray transitions from the decay of the ground state.
For each transition, the activity was determined as a function of time after
the end of irradiation and fit with an exponential decay function with the
half-life fixed to the evaluated value.
A weighted-average activity time series was then constructed from the four
individual transitions and fit with the same exponential form to obtain a
combined estimate of the initial activity.

\begin{figure}[t]
  \centering
  \includegraphics[width=1.10\columnwidth]{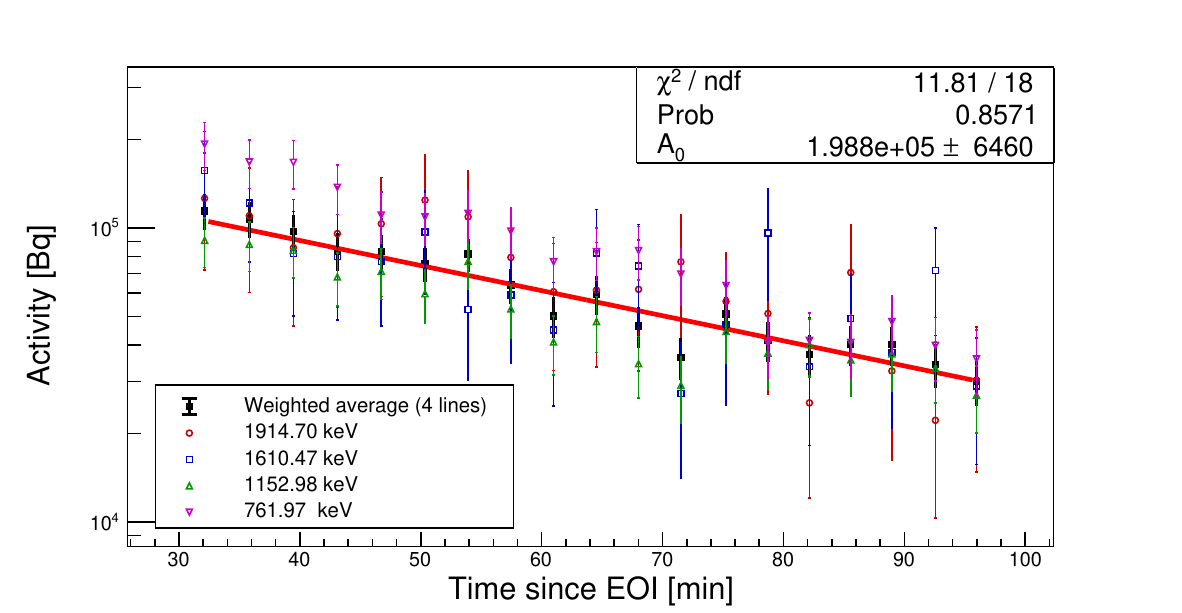}
  \caption{Measured activities of four $\gamma$-ray transitions from the decay
  of $^{111}$Sn as a function of time after EOI, together with their
  weighted-average activity. The solid curve shows an exponential fit to the
  weighted-average activity time series with the half-life fixed to the evaluated
  value, from which the initial activity is extracted. Individual fits to each
  $\gamma$-ray transition yield consistent initial activities.}
  \label{fig:sn111_weighted_A0}
\end{figure}

As a consistency check, each $\gamma$-ray time series was also fit independently,
and the resulting four initial activities were combined using a weighted average.
The initial activities obtained from the two approaches agree within their
respective uncertainties.
In the analysis presented here, we report the initial activity extracted from
the exponential fit to the weighted-average activity time series of the four
$\gamma$-ray transitions combined, as illustrated in
Fig.~\ref{fig:sn111_weighted_A0}.
An analogous procedure was applied to the remaining activation products.

For each measured reaction channel $r$ (corresponding to a unique reaction product), the calculated activity at EOI is modeled as
\begin{align}
\label{eq:A0_model}
A_{EOI,r}^{\mathrm{calc}}
&=
N_{T,k(r)}\,R_e\,\left(1-e^{-\lambda_r t_{\mathrm{irr}}}\right)\left(p_{\mathrm{lo}}\,S_{1,r}+p_{\mathrm{hi}}\,S_{2,r}\right),
\end{align}
where $k(r)$ denotes the target isotope associated with channel $r$, $N_{T,k(r)}$ is the corresponding 
number of target nuclei, and $R_e=I/e$ is the electron rate derived from the beam current $I$.

The coefficients $S_{1,r}$ and $S_{2,r}$ are band integrals over the low- and high-energy portions of 
the bremsstrahlung spectrum, scaled by the corresponding band-normalization parameters $p_{\mathrm{lo}}$ and $p_{\mathrm{hi}}$.

The band integrals $S_{1,r}$ and $S_{2,r}$ accumulate cross section $\times$ photon flux per electron 
over the two energy bands by partitioning the energy axis into sub-segments formed from the union of (i) 
MC flux-bin boundaries, (ii) cross-section tabulation boundaries, (iii) the band boundaries, and (iv) any explicit reaction threshold $E_{\mathrm{thr},r}$.
For band $\mathcal{B}_j$,
\begin{align}
\label{eq:band_sum}
S_{j,r}
&=
\sum_{\substack{\text{sub-segments } [x_0,x_1]\subset \mathcal{B}_j}}
\sigma_r(x_{0:1})\,
\Phi_{\text{sub}}([x_0,x_1]),
\end{align}
where $\Phi_{\text{sub}}([x_0,x_1])$ is the per-electron photon flux in that sub-segment 
(units $\mathrm{cm^{-2}\,e^{-1}}$), with $x_0$ and $x_1$ denoting the lower and upper energy 
bounds of each sub-segment formed by intersecting the photon-flux bins, cross-section tabulation 
bins, band limits, and the reaction threshold. The flux is obtained by splitting the parent MC bin 
proportionally to the segment width. The cross section $\sigma_r(x_{0:1})$ is treated as piecewise 
constant within the native library bins; since the sub-segmentation includes all tabulation edges, 
midpoint sampling is exact for piecewise-constant tabulations. Sub-segments with midpoint energy 
below $E_{\mathrm{thr},r}$ are excluded so that below-threshold contributions do not enter the sums.

\subsection{Propagation of flux uncertainties into the model variance}
\label{subsec:uncertainty}
\begin{figure}[t]
  \centering
  \includegraphics[width=1.10\columnwidth]{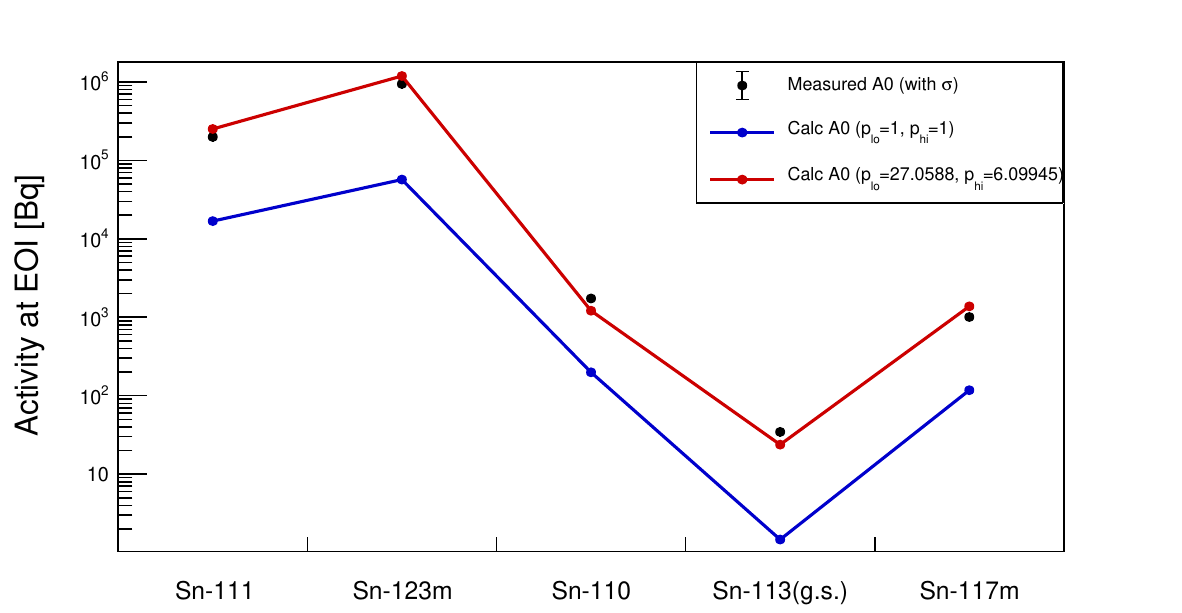}
  \caption{Comparison of measured initial activities ($A_{EOI}$) from the natural tin activation test
  with forward-model predictions using the uncorrected MC photon flux ($p_{\mathrm{lo}}=1$,
  $p_{\mathrm{hi}}=1$) and the corrected flux obtained from the $\chi^2$ minimization procedure.
  Error bars represent the reported experimental uncertainties (and propagated
  model/flux uncertainties used in the minimization). Lines connecting markers are included solely
  to improve visual identification of each data group and have no physical meaning.}
  \label{fig:tin_validation_A0}
\end{figure}

Each MC bin provides a relative flux uncertainty $\delta_b$ so that
$\mathrm{Var}[\Phi_b]=(\delta_b\,\Phi_b)^2$. Assuming uncorrelated bins, the band-variance contributions for reaction $r$ are

\begin{align}
V_{j,r}
&=
\sum_{\text{sub-intervals in }\mathcal{B}_j}
\left(
{\sigma}_r([x_0,x_1])
\right)^2
\,
\mathrm{Var}\!\left[
\Phi_{\text{sub}}([x_0,x_1])
\right],
\label{eq:band_variance}
\end{align}

where \(j\in\{\mathrm{lo},\mathrm{hi}\}\).

with $\mathrm{Var}[\Phi_{\text{sub}}]$ computed by scaling the parent bin variance by the fractional overlap area. 
The total prediction variance entering the $\chi^2$ for datum $r$ is
\begin{equation}
\mathrm{Var}[A_{EOI,r}^{\mathrm{calc}}] \;=\; \underbrace{\sigma_{A,r}^2}_{\text{measurement}}
\;+\;
\underbrace{C_r^2\left(p_{\mathrm{lo}}^2 V_{\mathrm{lo},r} + p_{\mathrm{hi}}^2 V_{\mathrm{hi},r}\right)}_{\text{flux model}}
\label{eq:model_var}
\end{equation}
This explicitly couples the fitted parameters to the model variance. Then, we determine $(p_{\mathrm{lo}},p_{\mathrm{hi}})$ by minimizing
\begin{equation}
\chi^2(p_{\mathrm{lo}},p_{\mathrm{hi}}) \;=\;
\sum_{r}
\frac{\big(A_{EOI,r}^{\mathrm{meas}}-A_{EOI,r}^{\mathrm{calc}}\big)^2}{
\sigma_{A,r}^2 + C_r^2\left(p_{\mathrm{lo}}^2 V_{\mathrm{lo},r} + p_{\mathrm{hi}}^2 V_{\mathrm{hi},r}\right)}
\end{equation}
using \textsc{Minuit}~\cite{BrunRademakers1997} minimizer to obtain best-fit values and covariance, with the number of degrees of 
freedom is $N_{\mathrm{data}}-2=3$. The fit is reported alongside measured vs.\ 
calculated activities. Figure~\ref{fig:tin_validation_A0} summarizes the tin-based flux validation by comparing the measured
initial activities with forward-model predictions obtained using (i) the uncorrected MC flux
($p_{\mathrm{lo}}=1$, $p_{\mathrm{hi}}=1$) and (ii) the corrected flux ($p_{\mathrm{lo}}=27.06$, $p_{\mathrm{hi}}=6.10$) derived from the $\chi^2$
minimization. 

Because the two normalization parameters rescale broad energy regions, the fitted values should be interpreted 
as effective spectral corrections accounting for converter-thickness–induced softening and residual geometry uncertainties. 
Therefore, these effective corrections absorb systematic effects arising from imperfections in the MC flux modeling and from uncertainties in
 the beam shape and absolute current determination at the low operating current of 1~\(\mu\)A.
 The corrected flux, expressed in Eq.~\ref{eq:corrected_flux}, is used in the extraction of the cross sections on the natural nickel target 
in Sec.~(\ref{sec:cross sections}). 
 
\section{Bremsstrahlung-averaged cross section analyses}
\label{sec:cross sections}
\begin{figure*}[t]
  \centering
  \includegraphics[width=\textwidth]{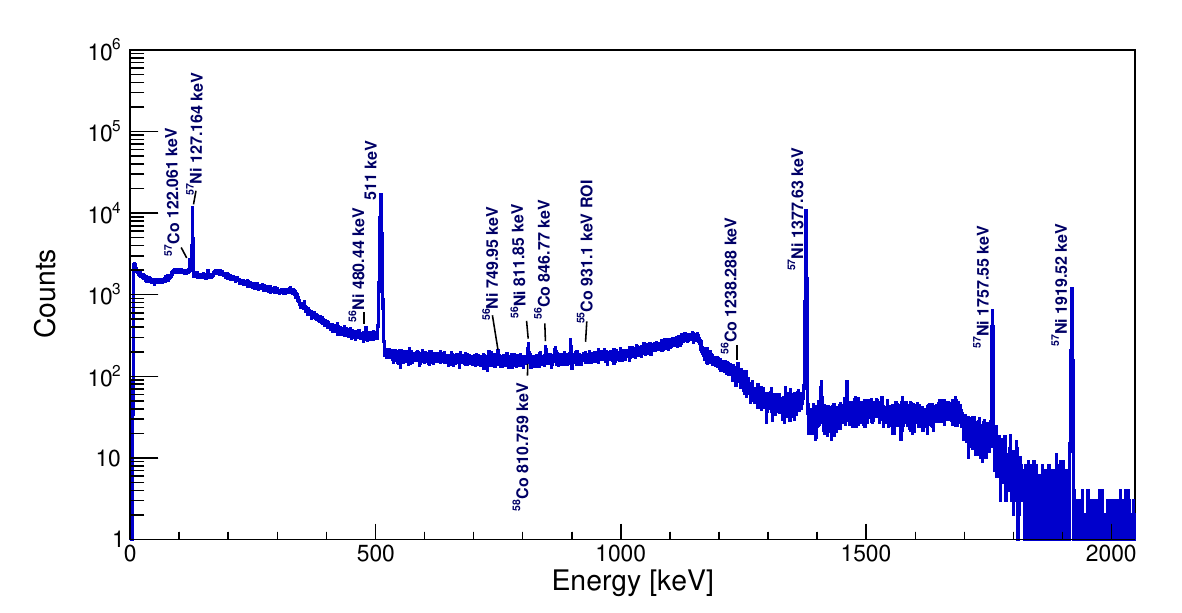}
  \caption{Energy-calibrated $\gamma$-ray spectrum acquired $\SI{1.37}{h}$ after the end of irradiation of the $^{nat}$Ni 
  target, shown over the full measured energy range. The spectrum corresponds 
  to a live counting time of $\SI{50400}{s}$ with a detector dead time of 0.22\%. Prominent photopeaks associated with the 
  decay of $^{57}$Ni, $^{56}$Ni, $^{58}$Co, $^{57}$Co, and $^{56}$Co are identified. The region of interest around the $^{55}$Co 
  $\gamma$-ray at 931.1~keV is also indicated.}
  \label{fig:ni_spectrum_full}
\end{figure*}

This section presents the determination of bremsstrahlung-averaged cross sections for photon-induced reactions on $^{nat}$Ni using the experimentally
validated bremsstrahlung photon flux. The analysis is based on the two-group
corrected MC flux for the \SI{4}{mm} (FWHM) beam spot case described in Sec.~\ref{sec:mc_model}, together with
measured initial activities extracted from post-irradiation $\gamma$-ray
spectroscopy. Six main reaction channels are considered in this work:
$^{58}\mathrm{Ni}(\gamma,n)^{57}\mathrm{Ni}$,
$^{58}\mathrm{Ni}(\gamma,2n)^{56}\mathrm{Ni}$,
$^{nat}\mathrm{Ni}(\gamma,pxn)^{58}\mathrm{Co}$,
$^{nat}\mathrm{Ni}(\gamma,pxn)^{57}\mathrm{Co}$,
$^{58}\mathrm{Ni}(\gamma,pn)^{56}\mathrm{Co}$, and
$^{58}\mathrm{Ni}(\gamma,p2n)^{55}\mathrm{Co}$. An overview of the $\gamma$-ray spectrum used to extract the initial activities is 
shown in Fig.~\ref{fig:ni_spectrum_full}. \\

\subsection{$^{nat}$Ni$(\gamma,xn)^{57}$Ni cross section analysis}

The production of $^{57}$Ni ($T_{1/2}=35.6$~h) in natural nickel targets irradiated with bremsstrahlung photons proceeds 
predominantly through the $^{58}$Ni$(\gamma,n)^{57}$Ni reaction. Natural nickel
 contains $^{58}$Ni with an isotopic abundance of 68.08\%, making this channel dominant above threshold. The reaction 
 threshold for $^{58}$Ni$(\gamma,n)$ is approximately $E_{\mathrm{thr}} \approx \SI{12.22}{\mega\electronvolt}$, 
with the cross section rising rapidly in the giant dipole resonance region between \SIrange{12.22}{40}{\mega\electronvolt}.
Although the higher-threshold reaction channel
$^{60}$Ni$(\gamma,3n)$ is energetically accessible at the present endpoint
energy it is estimated to contribute only 1.4\% of the measured activity yield, as discussed later in this section. 

Using the corrected bremsstrahlung flux, the bremsstrahlung-averaged cross section
for the dominant $^{58}$Ni$(\gamma,n)^{57}$Ni channel was extracted from the
measured EOI activity obtained via $\gamma$-ray spectroscopy.
The analysis employed three well-resolved $\gamma$-ray transitions from the decay
of $^{57}$Ni at $E_\gamma = \SI{1377.63}{keV}$, $\SI{1757.55}{keV}$ and \SI{1919.52}{keV}, with absolute
intensities of 0.817, 0.0575 and 0.123, respectively. For each transition, the net peak
area was corrected for detector efficiency and decay, yielding consistent EOI
activities within uncertainties. A weighted average of the three transitions was
used to determine the final $^{57}$Ni EOI activity, thereby minimizing
line-specific statistical and efficiency-related uncertainties.\\

The resulting measured bremsstrahlung-averaged cross section for
$^{58}$Ni$(\gamma,n)^{57}$Ni is

\begin{equation*}
\langle\sigma\rangle\!\left[^{58}\mathrm{Ni}(\gamma,n)^{57}\mathrm{Ni}\right]
= 8.983~\pm~0.028~\mathrm{mb}.
\end{equation*}

where the uncertainty reflects the combined contributions from
counting statistics, efficiency calibration, decay data, and the validated photon
flux normalization. This value is compared directly with the corresponding
JENDL-5–based bremsstrahlung-averaged cross section calculated using the same
corrected photon flux, which yields

\begin{equation*}
\langle\sigma\rangle_{\mathrm{JENDL-5}}\!\left[^{58}\mathrm{Ni}(\gamma,n)^{57}\mathrm{Ni}\right]
= \SI{10.293}{mb}.
\end{equation*}

The agreement between the measured and evaluated results is within the quoted uncertainties.

The contribution from the higher-threshold
$^{60}$Ni$(\gamma,3n)^{57}$Ni reaction was evaluated explicitly using the JENDL-5
cross section folded with the corrected bremsstrahlung spectrum and scaled by the
natural isotopic abundance of $^{60}$Ni. This channel contributes
approximately 1.4\% to the total $^{57}$Ni yield under the present irradiation
conditions, corresponding to an effective bremsstrahlung-averaged cross section
of $\langle\sigma\rangle \approx \SI{0.12}{mb}$. Given its small magnitude, this
contribution does not alter the extracted BACS within the experimental
uncertainties but is included for completeness in this evaluation.

\subsection{$^{58}$Ni$(\gamma,2n)^{56}$Ni cross section analysis}

In this irradiation of the natural nickel target, the formation of $^{56}$Ni ($T_{1/2}=6.081$~d) is
dominated purely by the $^{58}$Ni$(\gamma,2n)^{56}$Ni reaction. The threshold energy for the
$^{58}$Ni$(\gamma,2n)$ reaction is approximately
$E_{\mathrm{thr}} \approx \SI{21.8}{\mega\electronvolt}$, above which the cross
section increases across the giant dipole resonance region and remains
significant up to the bremsstrahlung endpoint energy. No contribution from higher-threshold channels, such as
$^{60}$Ni$(\gamma,4n)$ is expected, since they are energetically not allowed at the present endpoint.

The initial activity of $^{56}$Ni was determined using three $\gamma$-ray
transitions at $E_\gamma = \SI{480.44}{keV}$, \SI{749.95}{keV}, and
\SI{811.85}{keV}\nopagebreak\footnote{A spectral interference from the nearby $^{58}$Co ($T_{1/2}$=70.883~d) $\gamma$ line 
at \SI{810.76}{keV} ($I_{\gamma}$= 99.45\%) was resolved and is discussed in Sec.~\ref{sec:cross sections_Co58}.}.
 For each transition, the net peak area was corrected for
detector efficiency, branching intensity, and decay to obtain the
end-of-irradiation activity. The resulting activities are mutually
consistent within uncertainties, and a weighted average of the three transitions
was used to determine the final $^{56}$Ni EOI activity for the cross-section
extraction.

Using the corrected two-group bremsstrahlung photon flux, the measured
bremsstrahlung-averaged cross section for the
$^{58}$Ni$(\gamma,2n)^{56}$Ni reaction is
\begin{equation*}
\langle\sigma\rangle\!\left[^{58}\mathrm{Ni}(\gamma,2n)^{56}\mathrm{Ni}\right]
= 0.248~\pm~0.025~\mathrm{mb}.
\end{equation*}

The corresponding bremsstrahlung-averaged cross section obtained by folding the
JENDL-5 evaluated cross section with the same corrected photon flux is
\begin{equation*}
\langle\sigma\rangle_{\mathrm{JENDL-5}}\!\left[^{58}\mathrm{Ni}(\gamma,2n)^{56}\mathrm{Ni}\right]
= \SI{0.231}{mb}.
\end{equation*}

The measured result is consistent with the JENDL-5–based prediction within the
combined experimental and model uncertainties, indicating satisfactory
agreement between the evaluated photonuclear data and the present measurement.

\subsection{$^{nat}$Ni$(\gamma,pxn)^{58}$Co cross section analysis}
\label{sec:cross sections_Co58}

\begin{figure}[t]
  \centering
  \includegraphics[width=\columnwidth]{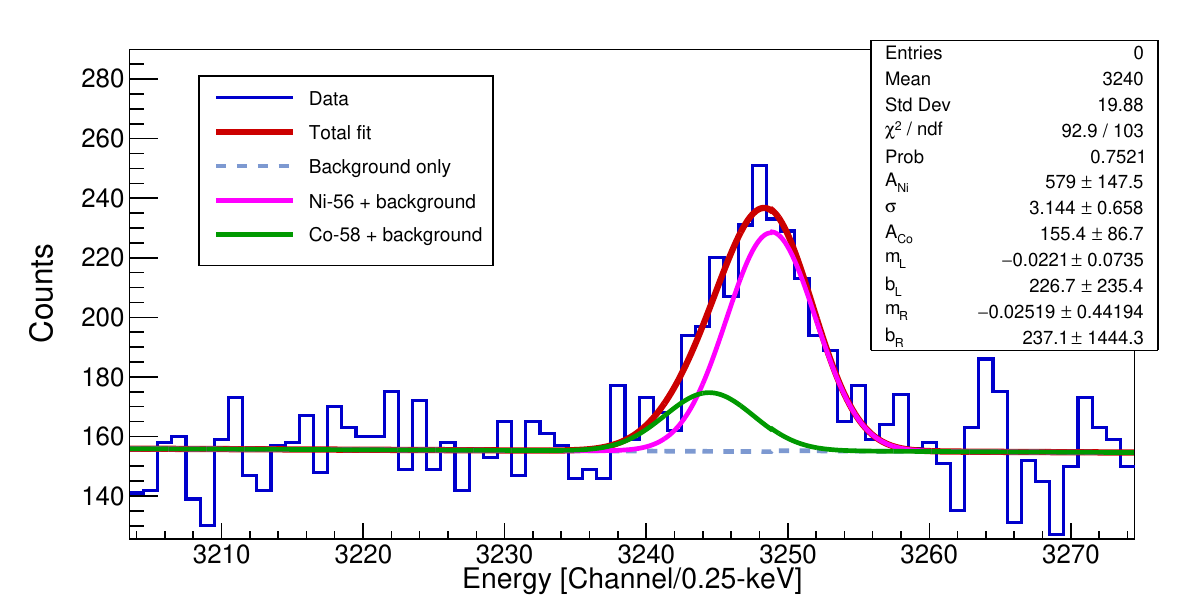}
  \caption{Section of the $\gamma$-ray spectrum showing the overlapping
  $^{58}\mathrm{Co}$ (\SI{810.76}{keV}) and $^{56}\mathrm{Ni}$ (\SI{811.85}{keV})
  photopeaks. The spectrum was fitted using a double-Gaussian function with a
  common width $\sigma$, superimposed on a linear background. The peak centroids
  were fixed to the channel values corresponding to the known $\gamma$-ray
  energies. The fit parameters, shown in the inset, include the left and right
  background intercepts ($b_L$, $b_R$), background slopes ($m_L$, $m_R$), peak
  areas ($A_{\mathrm{Co}}$, $A_{\mathrm{Ni}}$), and the shared Gaussian width
  $\sigma$. The extracted peak areas are $155.4~\pm~86.7$ counts for
  $^{58}\mathrm{Co}$ and $579~\pm~147.5$ counts for $^{56}\mathrm{Ni}$.}
  \label{fig:Co58}
\end{figure}

The production of $^{58}\mathrm{Co}$ proceeds through charged-particle emission channels collectively denoted as $^{nat}$Ni($\gamma,pxn)^{58}$Co. 
These channels open above the giant dipole resonance region and receive contributions from multiple stable nickel isotopes, 
with $^{60}$Ni providing the dominant yield component at the present bremsstrahlung end-point energy of \SI{40}{MeV}. 
Consequently, the extracted cross section represents an effective, flux-averaged quantity that incorporates all energetically 
accessible $(\gamma,pxn)$ pathways up to $x=3$.
Experimentally, the determination of the $^{58}$Co activity is complicated by the close proximity of its characteristic 
$\gamma$-ray to a transition from the decay of $^{56}$Ni. Figure~\ref{fig:Co58} illustrates the resulting spectral overlap 
and the procedure used to resolve the individual contributions from the two radionuclides.

Using the corrected two-group bremsstrahlung photon flux, the measured bremsstrahlung-averaged cross section for the 
$^{nat}$Ni($\gamma,pxn)^{58}$Co reaction is
\begin{equation*}
\langle\sigma\rangle\!\left[^{\mathrm{nat}}\mathrm{Ni}(\gamma,pxn)^{58}\mathrm{Co}\right]
= 0.704~\pm~0.218~\mathrm{mb}.
\end{equation*}

The corresponding bremsstrahlung-averaged cross section obtained by folding the JENDL-5 evaluated partial cross sections 
with the same corrected photon flux yields
\begin{equation*}
\langle\sigma\rangle_{\mathrm{JENDL\mbox{-}5}}\!\left[^{\mathrm{nat}}\mathrm{Ni}(\gamma,pxn)^{58}\mathrm{Co}\right]
= 0.216~\mathrm{mb}.
\end{equation*}

The experimental result is larger than the prediction based on JENDL-5, but remains within the experimental uncertainty, supporting the overall reliability of the evaluated photonuclear data for the production of $^{58}\mathrm{Co}$ 
from natural nickel under bremsstrahlung irradiation. 

Additionally, TALYS-2.2 calculations using the six available level density models indicate that only the default model ($L=1$) 
yields good agreement with the JENDL-5 result. The corresponding TALYS-2.2 bremsstrahlung-averaged cross sections for the $^{nat}$Ni($\gamma,pxn)^{58}$Co
 reaction are 0.183 mb ($L=1$), 0.028 mb ($L=2$), 0.038 mb ($L=3$), 
0.0025 mb ($L=4$), 0.029 mb ($L=5$), and 0.020 mb ($L=6$).

\subsection{Bremsstrahlung-averaged cross section for $^{57}$Co production}
\label{sec:cross_sections_Co57}

The direct production of $^{57}$Co under bremsstrahlung irradiation of a natural nickel
target is dominated by charged-particle emission reactions on $^{58}$Ni.
The primary contribution arises from the
\[
^{58}\mathrm{Ni}(\gamma,p)^{57}\mathrm{Co}
\]
reaction, which has a relatively low threshold energy of
$E_{\mathrm{thr}} = \SI{8.1729}{MeV}$ and is therefore strongly populated over
the full energy range of the bremsstrahlung photon spectrum up to the
\SI{40}{MeV} end-point energy used in the present experiment.

Additional contributions to the $^{57}$Co yield may originate from reactions
on heavier nickel isotopes. \\
In particular, the
\[
^{60}\mathrm{Ni}(\gamma,p2n)^{57}\mathrm{Co}
\]
channel opens at a substantially higher threshold energy,
$E_{\mathrm{thr}} = \SI{28.5662}{MeV}$, and contributes only a minor fraction
to the total yield due to both its higher reaction threshold and the natural
abundance of $^{60}$Ni ($26.2231\%$). Contributions from even heavier isotopes,
such as
\[
^{61}\mathrm{Ni}(\gamma,p3n)^{57}\mathrm{Co},
\]
are expected to be negligible at a \SI{40}{MeV} end-point energy, owing to their
high threshold energy ($E_{\mathrm{thr}} = \SI{36.3909}{MeV}$) and the low natural
abundance of $^{61}$Ni ($1.1399\%$). As a result, the measured bremsstrahlung-averaged
cross section for $^{57}$Co direct production is overwhelmingly governed by the
$^{58}$Ni$(\gamma,p)^{57}$Co reaction channel. However, there exists an indirect pathway for its production, which is via the 
formation and subsequent decay of $^{57}$Ni through
\[
^{58}\mathrm{Ni}(\gamma,n)^{57}\mathrm{Ni}
\;\xrightarrow[\;T_{1/2}=35.6~\mathrm{h}\;]{\varepsilon\;(100\%)}
\;
^{57}\mathrm{Co}.
\]
Although the half-life of $^{57}\mathrm{Ni}$ is long compared with the 1.37~h delay between irradiation and counting, 
it is not negligible over the 14~h counting interval. Consequently, the in-growth of $^{57}\mathrm{Co}$ from $^{57}\mathrm{Ni}$ decay 
must be explicitly accounted for, including decay during counting.

The total number of $^{57}\mathrm{Co}$ disintegrations, $N_{\mathrm{dis}}^{\mathrm{Co}}$, 
was determined from the spectral analysis after applying corrections for detector efficiency, dead time, and $\gamma$-ray emission probability. 
This quantity represents the combined contribution from directly produced 
$^{57}\mathrm{Co}$ and $^{57}\mathrm{Co}$ formed via $^{57}\mathrm{Ni}$ decay during the counting interval.

The observed number of disintegrations is related to the production mechanisms through the general expression

\begin{widetext}
\begin{equation}
\label{eq:Co57_general}
\begin{aligned}
N_{\mathrm{dis}}^{\mathrm{Co}}
&=
\int_{t_d}^{t_d+t_c}
\Bigl[
A_{\mathrm{Co}}^{\mathrm{dir}}(t)
+
A_{\mathrm{Co}}^{\mathrm{grow}}(t)
\Bigr] \, dt
\\[4pt]
&=
\frac{A_{\mathrm{Co}}^{\mathrm{dir}}(0)}{\lambda_{\mathrm{Co}}}
\left(
e^{-\lambda_{\mathrm{Co}} t_d}
-
e^{-\lambda_{\mathrm{Co}}(t_d+t_c)}
\right)
\\[4pt]
&\quad+
A_{^{57}\mathrm{Ni}}(0)\,
\frac{\lambda_{\mathrm{Co}}}{\lambda_{\mathrm{Co}}-\lambda_{\mathrm{Ni}}}
\left[
\frac{e^{-\lambda_{\mathrm{Ni}} t_d}-e^{-\lambda_{\mathrm{Ni}}(t_d+t_c)}}{\lambda_{\mathrm{Ni}}}
-
\frac{e^{-\lambda_{\mathrm{Co}} t_d}-e^{-\lambda_{\mathrm{Co}}(t_d+t_c)}}{\lambda_{\mathrm{Co}}}
\right]
\end{aligned}
\end{equation}
\end{widetext}

where $A_{\mathrm{Co}}^{\mathrm{dir}}(0)$ is the activity of $^{57}$Co at EOI produced directly via the $(\gamma,p)$ reaction, 
$A_{^{57}\mathrm{Ni}}(0)$ is the independently determined initial activity of $^{57}$Ni at EOI, and $\lambda_{\mathrm{Ni}}$ and 
$\lambda_{\mathrm{Co}}$ are the decay constants of $^{57}$Ni and $^{57}$Co, respectively.

Equation~\eqref{eq:Co57_general} fully accounts for the decay of directly produced $^{57}$Co as well as the time-dependent 
feeding from $^{57}$Ni during the entire counting interval. Solving this expression for the direct $^{57}$Co activity at EOI yields

\begin{widetext}
\begin{equation}
\label{eq:Co56_direct_activity}
A_{\mathrm{Co}}^{\mathrm{dir}}(0)
=
\lambda_{\mathrm{Co}}
\frac{
N_{\mathrm{dis}}^{\mathrm{Co}}
-
A_{^{57}\mathrm{Ni}}(0)\,
\frac{\lambda_{\mathrm{Co}}}{\lambda_{\mathrm{Co}}-\lambda_{\mathrm{Ni}}}
\left[
\frac{e^{-\lambda_{\mathrm{Ni}} t_d}-e^{-\lambda_{\mathrm{Ni}}(t_d+t_c)}}{\lambda_{\mathrm{Ni}}}
-
\frac{e^{-\lambda_{\mathrm{Co}} t_d}-e^{-\lambda_{\mathrm{Co}}(t_d+t_c)}}{\lambda_{\mathrm{Co}}}
\right]
}{
e^{-\lambda_{\mathrm{Co}} t_d}
-
e^{-\lambda_{\mathrm{Co}}(t_d+t_c)}
}
\end{equation}
\end{widetext}

The quantity $A_{\mathrm{Co}}^{\mathrm{dir}}(0)$ represents the initial activity at EOI of $^{57}$Co produced exclusively by 
the $^{58}$Ni$(\gamma,p)^{57}$Co reaction. Contributions from $^{60}$Ni$(\gamma,p2n)^{57}$Co are assumed negligible at a 40~MeV 
end-point energy and are therefore not considered further.

Using the corrected two-group bremsstrahlung photon flux, the flux-weighted
bremsstrahlung-averaged cross section for $^{57}$Co production in the present
work was determined to be
\begin{equation*}
\langle\sigma\rangle\!\left[^{\mathrm{58}}\mathrm{Ni}(\gamma,p)^{57}\mathrm{Co}\right]
=
9.192~\pm~0.386~\mathrm{mb},
\end{equation*}
where the quoted uncertainty reflects the combined statistical uncertainty of
the extracted $^{57}$Co activity and the propagated uncertainty associated with
the photon flux correction.

For comparison, the bremsstrahlung-averaged cross section obtained by folding the
JENDL-5 evaluated partial cross section for the dominant
$^{58}\mathrm{Ni}(\gamma,p)^{57}\mathrm{Co}$ channel with the same corrected photon
flux is
\begin{equation*}
\langle\sigma\rangle_{\mathrm{JENDL\mbox{-}5}}\!\left[^{\mathrm{58}}\mathrm{Ni}(\gamma,p)^{57}\mathrm{Co}\right]
=
4.600~\mathrm{mb}.
\end{equation*}
The measured cross section exceeds the JENDL-5 prediction by approximately a factor of two, indicating that the present results support a significantly larger effective contribution from the $(\gamma,p)$ reaction channel than is accounted for in the JENDL-5 evaluation under the same bremsstrahlung spectrum weighting.

In contrast, TALYS-2.2 calculations employing level density models $L=1$ through $L=6$ yield cross sections of 8.904 mb ($L=1$), 9.070 mb ($L=2$), 9.036 mb ($L=3$), 8.980 mb ($L=4$), 8.617 mb ($L=5$), and 8.668 mb ($L=6$). These values are in very good agreement with the experimental measurement reported in this work.

\subsection{Extraction of the $^{56}$Co bremsstrahlung-averaged cross section}

Similar to the case of $^{57}$Co, the production of $^{56}$Co proceeds through both direct and indirect reaction pathways. 
The direct contribution arises from the
\[
^{58}\mathrm{Ni}(\gamma,pn)^{56}\mathrm{Co}
\]
reaction, while an indirect contribution is produced via the formation and subsequent decay of $^{56}$Ni through
\[
^{58}\mathrm{Ni}(\gamma,2n)^{56}\mathrm{Ni}
\;\xrightarrow[\;T_{1/2}=6.081~\mathrm{d}\;]{\varepsilon\;(100\%)}
\;
^{56}\mathrm{Co}.
\]
Because the half-life of $^{56}$Ni is long compared to the irradiation-to-counting delay but not negligible over the counting 
interval, the feeding of $^{56}$Co from $^{56}$Ni must be treated explicitly, including decay during counting.

From the spectral analysis, the total number of $^{56}$Co disintegrations represents the sum of decays from both directly 
produced $^{56}$Co nuclei and those generated by $^{56}$Ni decay during the counting interval.

The observed number of disintegrations is related to the production mechanisms through the same general expression in 
Eq.~\ref{eq:Co57_general}, where $A_{\mathrm{Co}}^{\mathrm{dir}}(0)$ is the activity of $^{56}$Co at EOI produced directly 
via the $(\gamma,pn)$ reaction, $A_{^{56}\mathrm{Ni}}(0)$ is the independently determined initial activity of $^{56}$Ni at 
EOI, and $\lambda_{\mathrm{Ni}}$ and $\lambda_{\mathrm{Co}}$ are the decay constants of $^{56}$Ni and $^{56}$Co, respectively.

The direct $^{56}$Co activity at EOI is calculated using

\begin{widetext}
\begin{equation}
\label{eq:Co56_direct_activity}
A_{\mathrm{Co}}^{\mathrm{dir}}(0)
=
\lambda_{\mathrm{Co}}
\frac{
N_{\mathrm{dis}}^{\mathrm{Co}}
-
A_{^{56}\mathrm{Ni}}(0)\,
\frac{\lambda_{\mathrm{Co}}}{\lambda_{\mathrm{Co}}-\lambda_{\mathrm{Ni}}}
\left[
\frac{e^{-\lambda_{\mathrm{Ni}} t_d}-e^{-\lambda_{\mathrm{Ni}}(t_d+t_c)}}{\lambda_{\mathrm{Ni}}}
-
\frac{e^{-\lambda_{\mathrm{Co}} t_d}-e^{-\lambda_{\mathrm{Co}}(t_d+t_c)}}{\lambda_{\mathrm{Co}}}
\right]
}{
e^{-\lambda_{\mathrm{Co}} t_d}
-
e^{-\lambda_{\mathrm{Co}}(t_d+t_c)}
}
\end{equation}
\end{widetext}

The quantity $A_{\mathrm{Co}}^{\mathrm{dir}}(0)$ represents the initial activity at EOI of $^{56}$Co produced exclusively 
by the $^{58}$Ni$(\gamma,pn)^{56}$Co reaction. Contributions from $^{60}$Ni$(\gamma,p3n)^{56}$Co are energetically 
forbidden at a 40~MeV end-point energy and are therefore not considered further.

The bremsstrahlung-averaged cross section for the $^{58}$Ni$(\gamma,pn)^{56}$Co reaction is then obtained using
\begin{equation}
\label{eq:Co56_bacs}
\langle \sigma \rangle_{^{58}\mathrm{Ni}(\gamma,pn)^{56}\mathrm{Co}}
=
\frac{
A_{\mathrm{Co}}^{\mathrm{dir}}(0)
}{
N_{58}\,\Phi^{\mathrm{corr}}(E_{\mathrm{thr}},40)\,
\left(1-e^{-\lambda_{\mathrm{Co}} t_{\mathrm{irr}}}\right)
}
\end{equation}
where $N_{58}$ is the number of $^{58}$Ni target atoms, and $\Phi^{\mathrm{corr}}(E_{\mathrm{thr}},40)$ is the 
corrected bremsstrahlung photon flux integral from the reaction threshold energy $E_{\mathrm{thr}}$ up to the 40~MeV 
end-point energy.

Using the two-group energy-dependent cross section matched to the corrected two-group bremsstrahlung photon flux binning, 
the bremsstrahlung-averaged cross section for the $^{58}$Ni$(\gamma,pn)^{56}$Co reaction was determined to be
\[
\langle \sigma \rangle\left[^{58}\mathrm{Ni}(\gamma,pn)^{56}\mathrm{Co}\right] = 2.239~\pm~0.355~\mathrm{mb},
\]
where the quoted uncertainty reflects the combined statistical uncertainty of the measured $^{56}$Co disintegrations and 
the propagated uncertainty from the $^{56}$Ni feeding correction. This result is about a factor of two higher than the evaluated JENDL-5 
bremsstrahlung-averaged cross section of
\[
\langle \sigma \rangle_{\mathrm{JENDL-5}}\left[^{58}\mathrm{Ni}(\gamma,pn)^{56}\mathrm{Co}\right] = 1.179~\mathrm{mb},
\]
with the difference attributed primarily to uncertainties in the high-energy tail of the bremsstrahlung photon 
flux and the energy dependence of the $(\gamma,pn)$ reaction near threshold. 

A comparison with TALYS-2.2 calculations (level density models $L=1$ through $L=6$) shows reasonable agreement with JENDL-5, 
with predicted bremsstrahlung-averaged cross sections of 1.522 mb ($L=1$), 1.337 mb ($L=2$), 1.474 mb ($L=3$), 
1.538 mb ($L=4$), 1.649 mb ($L=5$), and 1.334 mb ($L=6$) for the $^{58}$Ni$(\gamma,pn)^{56}$Co reaction.

\subsection{$^{55}$Co cross section analysis}

\begin{figure}[t]
  \centering
  \includegraphics[width=1.0\linewidth]{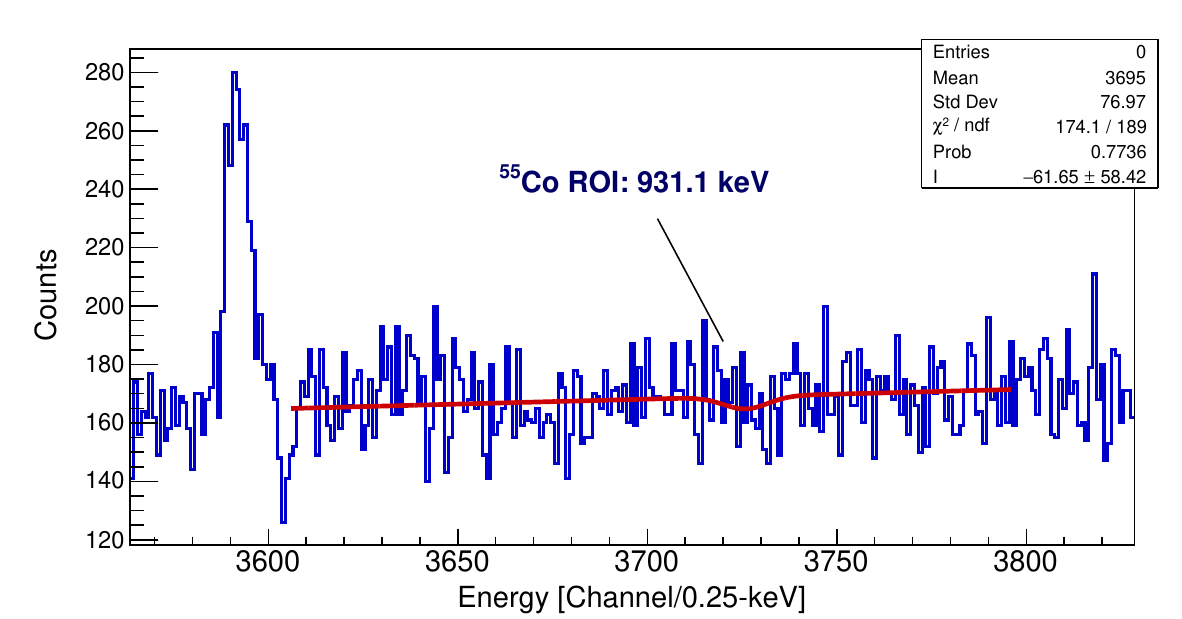}
  \caption{The ROI used for the determination of the upper limit on the initial activity of \textsuperscript{55}Co from 
  the 931.1~keV $\gamma$-ray transition ($I_\gamma = 75\%$). 
  A linear background in the vicinity of the peak ($\pm100$ channels around the ROI) was fitted independently, and the 
  resulting background parameters were subsequently fixed in the (Gaussian + background) peak fit. 
  In the final fit, the centroid was fixed at channel 3725.97 based on the energy calibration, while the Gaussian standard 
  deviation was fixed to 5.81 channels, as extracted from Gaussian fits to nearby well-resolved peaks. 
  The detector full-energy peak efficiency at 931.1~keV was taken as $3.00\times10^{-5}$. 
  This constrained fitting procedure is used to determine a one-sided 90\% C.L. upper limit on the 
  initial \textsuperscript{55}Co activity where the net area under the peak is determined from the fit to be -61.65~$\pm$ 58.42 counts.}
  \label{fig:Co55_ROI}
\end{figure}

In cases where the extracted bremsstrahlung-averaged cross section is consistent with zero within statistical uncertainty, 
or where the fitted yield (equivalently, the measured net area under the photopeak, see Fig.~\ref{fig:Co55_ROI}) is negative 
and therefore non-physical, a one-sided confidence interval is adopted to determine an upper limit. In this work, a one-sided 
Gaussian construction is used to report a 90\% confidence level (C.L.) upper limit on the bremsstrahlung-averaged cross section.
As illustrated in Fig.~\ref{fig:Co55_ROI}, the upper limit is derived from the ROI surrounding the 931.1~keV $\gamma$-ray 
transition of \textsuperscript{55}Co. The procedure accounts for the fitted background and detector response parameters, 
which are constrained independently to ensure a physically meaningful and statistically robust determination of the upper 
limit. Under this prescription, the upper limit is defined as
\begin{equation}
\langle \sigma \rangle_{\mathrm{UL}} = \hat{\sigma} + 1.645\,\delta\hat{\sigma},
\end{equation}
where $\hat{\sigma}$ is the fitted bremsstrahlung-averaged cross section and $\delta\hat{\sigma}$ is its total uncertainty. 
The factor of 1.645 corresponds to the one-sided 90\% quantile of the standard normal distribution. 

Using this method, an upper limit of $\langle \sigma \rangle <0.021~\mathrm{mb}$ at the 90\% C.L. is obtained for the 
reaction channel of interest. This upper limit is interpreted as arising purely from photon-induced reactions on $^{58}$Ni 
via the $(\gamma,p2n)$ channel. The reaction threshold for $^{58}\mathrm{Ni}(\gamma,p2n)^{55}\mathrm{Co}$ 
is $E_{\mathrm{thr}} = 22.9639~\mathrm{MeV}$, which is well below the endpoint energy of the 40~MeV bremsstrahlung 
beam. Consequently, this channel is energetically accessible over an appreciable portion of the bremsstrahlung 
spectrum and represents the only physically allowed production mechanism contributing to the observed limit 
under the present irradiation conditions.

In contrast, the competing $^{58}\mathrm{Ni}(\gamma,3n)^{55}\mathrm{Ni}$ reaction, followed by electron capture
 decay of $^{55}$Ni with a half-life of 203.9~ms, is not energetically accessible for the majority of the 
 bremsstrahlung spectrum. The threshold energy for the $(\gamma,3n)$ reaction is $E_{\mathrm{thr}} = 39.121~\mathrm{MeV}$, which
 lies very near the bremsstrahlung endpoint energy of 40~MeV. As a result, only an infinitesimally small fraction of the photon 
 flux is capable of inducing this reaction. This behavior is reflected in the evaluated partial cross sections, which show a
 clear onset only at energies approaching the endpoint, with negligible contribution below 40~MeV.

Given the sharply suppressed photon flux near the bremsstrahlung endpoint and the near-threshold nature of the $(\gamma,3n)$
 channel, its contribution to the bremsstrahlung-averaged cross section is expected to be negligible. Therefore, the reported 90\% C.L. upper limit of:
\[
\langle \sigma \rangle\left[^{58}\mathrm{Ni}(\gamma,p2n)^{55}\mathrm{Co}\right] < 0.021~\mathrm{mb},
\]
can be attributed exclusively to the $^{58}\mathrm{Ni}(\gamma,p2n)$ reaction channel. This interpretation is consistent 
with both reaction threshold considerations and the observed energy dependence of the relevant partial cross sections.
However, this result shows a disagreement with the evaluated JENDL-5 bremsstrahlung-averaged cross section of:
\[
\langle \sigma \rangle_{\mathrm{JENDL-5}}\left[^{58}\mathrm{Ni}(\gamma,p2n)^{55}\mathrm{Co}\right] = 0.035~\mathrm{mb}.
\]

\section{Comparison with Previous Bremsstrahlung-Averaged Measurements}
\label{sec:lit_comparison}

Although this work and prior studies~\cite{Deiev2022NiPhotonuclear,Timchenko2025Co5765Photonuclear,Timchenko2025Co58Photonuclear} 
operated at nominally the same endpoint energy, the reported bremsstrahlung-averaged cross sections differ slightly. Table~\ref{tab:bacs_compare}
 summarizes representative values for the dominant nickel channels. The contrast originates from different photon spectra at the target, driven by
 converter thickness and downstream filtering.

\begin{table}[t]
\centering
\caption{Comparison of bremsstrahlung-averaged cross sections measured in this work (Ta converter thickness \SI{6}{\milli\meter}, no Al absorber)
 with literature values~\cite{Deiev2022NiPhotonuclear,Timchenko2025Co5765Photonuclear,Timchenko2025Co58Photonuclear} obtained using
 a 40.10~MeV electron beam, a thin Ta converter (\SI{1.05}{\milli\meter}), and a \SI{15}{\centi\meter} Al absorber. Uncertainties
 correspond to one standard deviation. The literature values represent averages of cross sections independently extracted from two
 $\gamma$-ray transitions for each reaction product.}
\label{tab:bacs_compare}
\begin{tabular}{@{}lclc@{}}
\toprule
Reaction & BACS (mb)      & BACS (mb) \\
         & (this work)    &  (literature)\\
\midrule
$^{58}$Ni$(\gamma,n)\,^{57}$Ni   & $8.983~\pm~0.028$ & $6.945~\pm~1.191$~\cite{Deiev2022NiPhotonuclear} \\
$^{58}$Ni$(\gamma,2n)\,^{56}$Ni & $0.248~\pm~0.025$ & $0.350~\pm~0.044$~\cite{Deiev2022NiPhotonuclear}\\
$^{nat}$Ni($\gamma,pxn)^{58}$Co & $0.704~\pm~0.218$  & $1.35~\pm~0.12$ ~\cite{Timchenko2025Co58Photonuclear}\\
$^{58}$Ni($\gamma,p)^{57}$Co & $9.192~\pm~0.386$  & $25.5~\pm~2.1$ ~\cite{Timchenko2025Co5765Photonuclear}\\
$^{58}$Ni($\gamma,pn)^{56}$Co & $2.239~\pm~0.355$  & $4.31~\pm~0.30$ ~\cite{Timchenko2025Co5765Photonuclear}\\
$^{58}$Ni($\gamma,p2n)^{55}$Co & $<0.021$  & $0.148~\pm~0.018$ ~\cite{Timchenko2025Co5765Photonuclear}\\
\bottomrule
\end{tabular}
\end{table}

\begin{figure*}[p]
\centering
\captionsetup{justification=raggedright,singlelinecheck=false}

% --- Top row: (a,b)
\begin{subfigure}{\linewidth}
  \centering
\includegraphics[width=\textwidth,height=0.3\textheight,keepaspectratio]{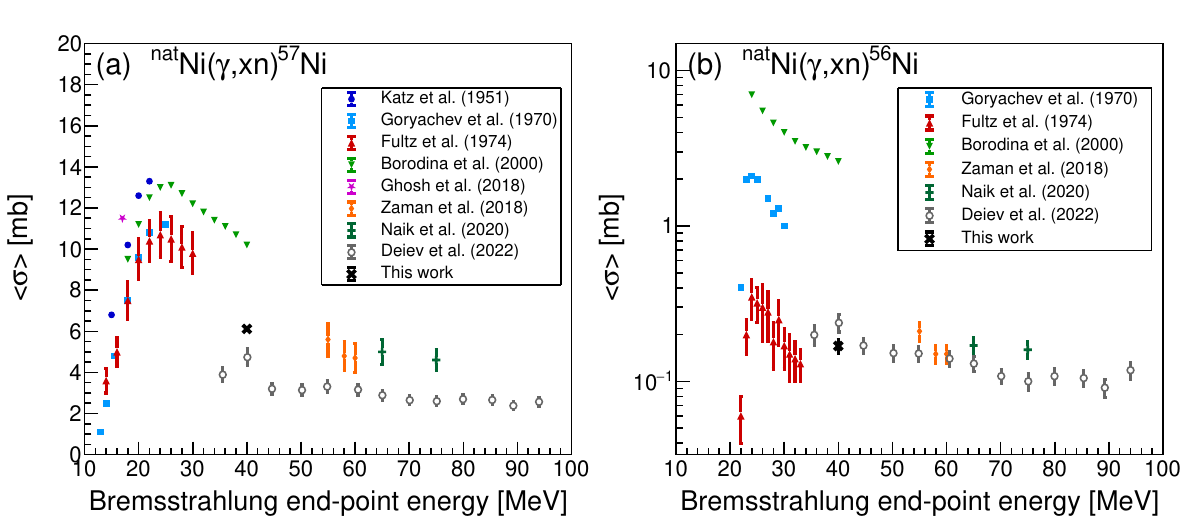}
\end{subfigure}

\vspace{0.6em}

% --- Middle row: (c,d)
\begin{subfigure}{\linewidth}
  \centering
\includegraphics[width=\textwidth,height=0.3\textheight,keepaspectratio]{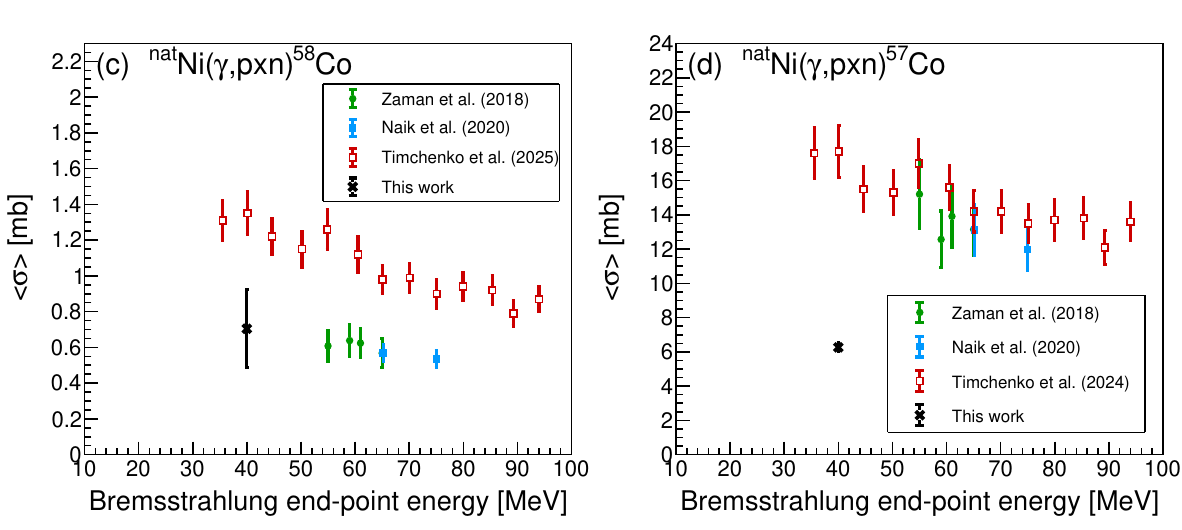}

\end{subfigure}

\vspace{0.6em}

% --- Bottom row: (e,f)  (force SAME width as others)
\begin{subfigure}{\linewidth}
  \centering
\includegraphics[width=\textwidth,height=0.3\textheight,keepaspectratio]{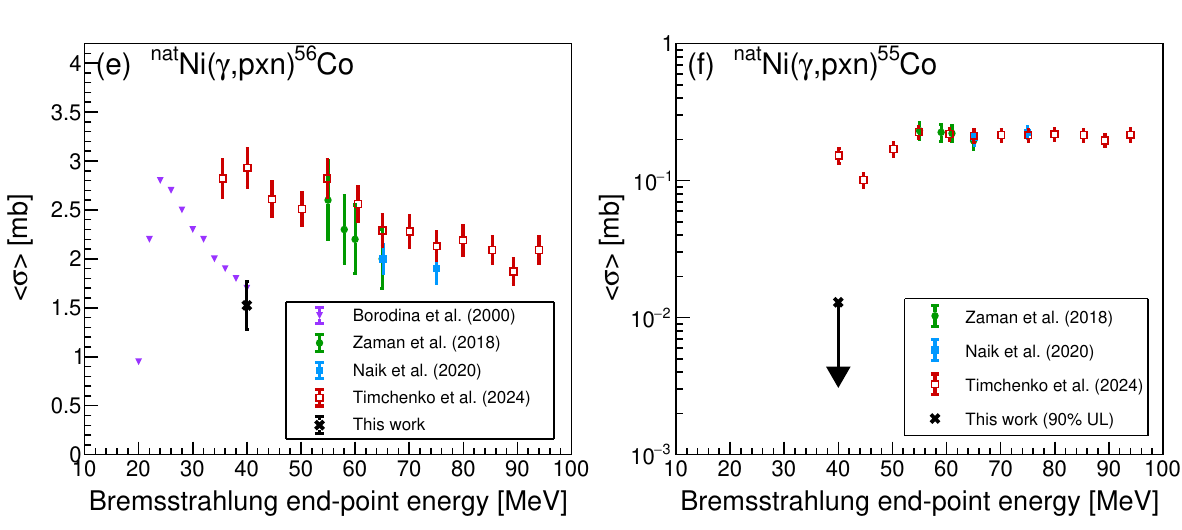}
\end{subfigure}

\caption{%
Bremsstrahlung-averaged photonuclear cross sections
$\langle\sigma\rangle$ for reactions on a natural nickel target as a function of
the bremsstrahlung end-point energy.
The panels show: (a) $^{\mathrm{nat}}\mathrm{Ni}(\gamma,xn)^{57}\mathrm{Ni}$,
(b) $^{\mathrm{nat}}\mathrm{Ni}(\gamma,xn)^{56}\mathrm{Ni}$,
(c) $^{\mathrm{nat}}\mathrm{Ni}(\gamma,pxn)^{58}\mathrm{Co}$,
(d) $^{\mathrm{nat}}\mathrm{Ni}(\gamma,pxn)^{57}\mathrm{Co}$,
(e) $^{\mathrm{nat}}\mathrm{Ni}(\gamma,pxn)^{56}\mathrm{Co}$, and
(f) $^{\mathrm{nat}}\mathrm{Ni}(\gamma,pxn)^{55}\mathrm{Co}$.
Literature datasets plotted here were digitized from Figs.~3--5 and Table~3 of
Ref.~\cite{Naik2020NatNiCo58}, Table~2 of Ref.~\cite{Deiev2022NiPhotonuclear}, and Tables~2-4 of Ref.~\cite{Timchenko2025Co5765Photonuclear}, for
$^{\mathrm{nat}}\mathrm{Ni}(\gamma,pxn)^{58}\mathrm{Co}$, supplemented by
Ref.~\cite{Timchenko2025Co58Photonuclear}; they are shown alongside the present measurements.}
\label{fig:natNi_BACS}
\end{figure*}

One factor that warrants particular consideration in explaining the minor discrepancies with published results is the converter
 thickness and its influence on the photon spectral hardness. The irradiation configuration of~\cite{Deiev2022NiPhotonuclear} and
 \cite{Timchenko2025Co58Photonuclear} used a thin Ta converter (\SI{1.05}{\milli\meter}), which produces a spectrum closer to the
 ideal thin-target (Bethe–Heitler-like~\cite{BetheHeitler1934Stopping}) shape. By contrast, the present work employed a \SI{6}{\milli\meter} Ta
 converter ($t/R_{\mathrm{CSDA}}\!\approx\!0.73$ at \SI{40}{MeV}), where electrons radiate repeatedly while slowing down. The result is a softer
 spectrum with enhanced photon fluence below $\sim\,$\SI{15}{MeV} and a depleted high-energy tail at the converter exit. This redistribution
 of spectral weight changes the folding integral that defines the BACS.\\
A broader comparison with previously published measurements is shown in
Fig.~\ref{fig:natNi_BACS}, which compiles bremsstrahlung-averaged cross sections
for nickel and cobalt production channels on a natural nickel target over a wide
range of end-point energies. Panels (a) and (b) compare the present
$^{\mathrm{nat}}$Ni$(\gamma,xn)^{57,56}$Ni results with early measurements by
Katz \emph{et al.}~\cite{Katz1951}, Goryachev \emph{et al.}~\cite{Goryachev1970},
Fultz \emph{et al.}~\cite{Fultz1974NiPhotonuclear}, Borodina \emph{et al.}~\cite{Borodina2000},
Ghosh \emph{et al.}~\cite{Ghosh2018}, Zaman \emph{et al.}~\cite{Zaman2018NatNiPhotonuclear},
Naik \emph{et al.}~\cite{Naik2020NatNiCo58}, and Deiev \emph{et al.}~\cite{Deiev2022NiPhotonuclear}.
Panels (c)–(f) present the corresponding cobalt production channels,
including $^{58}$Co, $^{57}$Co, $^{56}$Co, and $^{55}$Co, together with data from
Zaman \emph{et al.}~\cite{Zaman2018NatNiPhotonuclear}, Naik \emph{et al.}~\cite{Naik2020NatNiCo58},
Borodina \emph{et al.}~\cite{Borodina2000}, Timchenko \emph{et al.}~\cite{Timchenko2025Co5765Photonuclear}, and the
 recent $^{58}$Co measurements of Timchenko \emph{et al.}~\cite{Timchenko2025Co58Photonuclear}.

For consistency with these literature datasets, the bremsstrahlung-averaged cross
sections reported in the present work were adjusted to reflect the same effective
isotopic normalization adopted by the cited authors. Specifically, the measured
values for the $^{57}$Ni and $^{56}$Ni production channels shown in
Fig.~\ref{fig:natNi_BACS}(a,b), as well as the cobalt channels in
Fig.~\ref{fig:natNi_BACS}(d)–(f), were multiplied by the natural isotopic abundance
of $^{58}$Ni, while the $^{58}$Co channel in Fig.~\ref{fig:natNi_BACS}(c) was scaled
by the natural abundance of $^{60}$Ni. This procedure converts the yields obtained
from a natural nickel target into isotope-normalized cross sections that are
directly comparable to measurements and evaluations reported in the literature.
After applying this normalization, the remaining differences between datasets can
be attributed primarily to variations in the bremsstrahlung spectral shape and
experimental conditions rather than to isotopic composition effects.

\section{Conclusions}
\label{sec:conclusion}
Bremsstrahlung-averaged cross sections have been measured for photon-induced
reactions on a natural nickel target irradiated with a \SI{40}{MeV} endpoint
bremsstrahlung photon field. The analysis combined high-resolution $\gamma$-ray
spectroscopy with a validated MC description of the bremsstrahlung
photon flux, enabling the extraction of reliable flux-averaged observables under
production-relevant irradiation conditions.

A central component of this work was the experimental validation of the
MC photon-flux model using activation measurements on a natural tin
target. By constraining a two-group flux parameterization through a
$\chi^2$ minimization procedure applied to five independent tin activation
products, the photon flux was adjusted in a manner that explicitly accounts for
both experimental and model uncertainties. The validated flux was subsequently
used without further tuning in the analysis of the nickel irradiations, thereby
decoupling flux determination from cross-section extraction.

Using this approach, bremsstrahlung-averaged cross sections were determined for
the reactions
$^{58}$Ni$(\gamma,n)^{57}$Ni,
$^{58}$Ni$(\gamma,2n)^{56}$Ni, 
$^{nat}$Ni$(\gamma,pxn)^{58}$Co, $^{58}$Ni$(\gamma,p)^{57}$Co, and
$^{58}$Ni$(\gamma,pn)^{56}$Co.
For $^{57}$Ni, $^{56}$Ni and $^{58}$Co, the extracted cross sections are in good agreement
with JENDL-5 evaluated values folded with the same validated photon spectrum,
indicating consistency between the present measurements and modern photonuclear
evaluations when spectral effects are properly accounted for. This consistency is
further supported by TALYS-2.2 calculations, which reproduce the evaluated trends
for these channels when appropriate model parameters are used. In the case of
$^{56}$Co, the analysis explicitly treated the time-dependent feeding from
$^{56}$Ni decay during the counting interval, allowing the direct
$(\gamma,pn)$ contribution to be isolated and quantified. For both the
$^{58}$Ni$(\gamma,p)^{57}$Co and $^{58}$Ni$(\gamma,pn)^{56}$Co channels, TALYS-2.2
predictions are in reasonable agreement with JENDL-5, while the present measurements
tend to be higher, by up to approximately a factor of two, reflecting enhanced
charged-particle emission strengths and remaining sensitivities to model inputs
and bremsstrahlung spectral uncertainties.

For the $^{58}$Ni$(\gamma,p2n)^{55}$Co reaction, no statistically significant
signal was observed. Hence, a one-sided 90\% confidence-level upper limit on the
bremsstrahlung-averaged cross section was established. Reaction
threshold considerations and evaluated cross sections indicate that this limit
can be attributed exclusively to the $(\gamma,p2n)$ channel, with competing
higher-multiplicity neutron-emission reactions being strongly suppressed by the
steeply falling bremsstrahlung spectrum near the endpoint energy.

Comparison with previously published measurements performed at nominally the same bremsstrahlung endpoint energy reveals 
systematic differences that can be attributed primarily to variations in converter thickness and the resulting photon 
spectral hardness. In particular, the use of a thick tantalum converter in the present work produces a softer bremsstrahlung 
spectrum relative to the thin‑converter configurations employed in earlier studies, thereby altering the energy weighting of 
the photonuclear cross sections entering the bremsstrahlung‑averaging integral. Although the present measurements were performed 
at a single endpoint energy, the combination of an experimentally validated photon spectrum and explicit comparison with datasets 
obtained under different converter configurations allows the energy‑dependent sensitivity of bremsstrahlung‑averaged observables 
to be assessed indirectly. In this sense, the present results provide not only new integral cross‑section data but also a 
benchmark framework for interpreting bremsstrahlung‑averaged measurements obtained under non‑identical spectral and 
geometric conditions across different facilities.

Overall, this work demonstrates a quantitatively constrained methodology for extracting bremsstrahlung-averaged 
photonuclear cross sections from activation measurements. The combined use of experimentally validated photon fluxes, 
explicit uncertainty propagation, and careful treatment of decay feeding yields an approach that reflects realistic 
production conditions and is readily extendable to other target materials and reaction channels of relevance to 
medical isotope production and photonuclear data benchmarking. It is important to emphasize that the present work 
is not intended as an energy-dependent survey measurement, but rather as a precision benchmark at a fixed, 
production-relevant endpoint energy. The value of this study lies in the integration of spectral validation, rigorous 
uncertainty treatment, and representative irradiation conditions, enabling a more robust assessment of 
bremsstrahlung-averaged cross sections than is typically achievable in multi-energy survey measurements reported 
in the literature.

\section{Acknowledgments}
\label{sec:Acknowledgment}
The authors thank NMR for granting access to the accelerator and beamline, as well as the experimental data
 and laboratory facilities that made this work possible. We also acknowledge Benjamin Puffer, Jacquelyn Duty,
 and Robert Gotz for their valuable support with nickel target preparation and assistance during data acquisition.
%% ---------------- Bibliography ----------------
%\bibliographystyle{elsarticle-harv}
\bibliographystyle{unsrturl}
\bibliography{refs}

\begin{thebibliography}{10}

\bibitem{Brown2018}
D.~A. Brown, M.~B. Chadwick, R.~Capote, A.~C. Kahler, A.~Trkov, M.~W. Herman,
  A.~A. Sonzogni, Y.~Danon, A.~D. Carlson, M.~Dunn, D.~L. Smith, G.~M. Hale,
  G.~Arbanas, R.~Arcilla, C.~R. Bates, B.~Beck, B.~Becker, F.~Brown, R.~J.
  Casperson, J.~Conlin, D.~E. Cullen, M.-A. Descalle, R.~Firestone, T.~Gaines,
  K.~H. Guber, A.~I. Hawari, J.~Holmes, T.~D. Johnson, T.~Kawano, B.~C.
  Kiedrowski, A.~J. Koning, S.~Kopecky, L.~Leal, J.~P. Lestone, C.~Lubitz,
  J.~I. Márquez~Damián, C.~M. Mattoon, E.~A. McCutchan, S.~Mughabghab,
  P.~Navrátil, D.~Neudecker, G.~P.~A. Nobre, G.~Noguère, M.~Paris, M.~T.
  Pigni, A.~J. Plompen, B.~Pritychenko, V.~G. Pronyaev, D.~Roubtsov,
  D.~Rochman, P.~Romano, P.~Schillebeeckx, S.~Simakov, M.~Sin, I.~Sirakov,
  B.~Sleaford, V.~Sobes, E.~S. Soukhovitskii, I.~Stetcu, P.~Talou, I.~Thompson,
  S.~van~der Marck, L.~Welser-Sherrill, D.~Wiarda, M.~White, J.~L. Wormald,
  R.~Q. Wright, M.~Zerkle, G.~Žerovnik, and Y.~Zhu.
\newblock {ENDF/B-VIII.0}: The 8th major release of the nuclear reaction data
  library with {CIELO}-project cross sections, new standards and thermal
  scattering data.
\newblock {\em Nuclear Data Sheets}, 148:1--142, 2018.
\newblock \href {https://doi.org/10.1016/j.nds.2018.02.001}
  {\path{doi:10.1016/j.nds.2018.02.001}}.

\bibitem{Koning2019}
A.~J. Koning, D.~Rochman, J.-Ch. Sublet, N.~Dzysiuk, M.~Fleming, and S.~van~der
  Marck.
\newblock {TENDL}: complete nuclear data library for innovative nuclear science
  and technology.
\newblock {\em Nuclear Data Sheets}, 155:1--55, 2019.
\newblock \href {https://doi.org/10.1016/j.nds.2019.01.002}
  {\path{doi:10.1016/j.nds.2019.01.002}}.

\bibitem{Kawano2020}
T.~Kawano, Y.~S. Cho, P.~Dimitriou, D.~Filipescu, N.~Iwamoto, V.~Plujko,
  X.~Tao, H.~Utsunomiya, V.~Varlamov, R.~Xu, R.~Capote, I.~Gheorghe,
  O.~Gorbachenko, Y.L. Jin, T.~Renstr{\o}m, M.~Sin, K.~Stopani, Y.~Tian, G.~M.
  Tveten, J.~M. Wang, T.~Belgya, R.~Firestone, S.~Goriely, J.~Kopecky,
  M.~Krti{\v c}ka, R.~Schwengner, S.~Siem, and M.~Wiedeking.
\newblock Iaea photonuclear data library 2019.
\newblock {\em Nuclear Data Sheets}, 163:109--162, 2020.
\newblock \href {https://doi.org/10.1016/j.nds.2019.12.002}
  {\path{doi:10.1016/j.nds.2019.12.002}}.

\bibitem{Iwamoto2023}
Osamu Iwamoto, Nobuyuki Iwamoto, Satoshi Kunieda, Futoshi Minato, Shinsuke
  Nakayama, Yutaka Abe, Kohsuke Tsubakihara, Shin Okumura, Chikako Ishizuka,
  Tadashi Yoshida, Satoshi Chiba, Naohiko Otuka, Jean-Christophe Sublet, Hiroki
  Iwamoto, Kazuyoshi Yamamoto, Yasunobu Nagaya, Kenichi Tada, Chikara Konno,
  Norihiro Matsuda, Kenji Yokoyama, Hiroshi Taninaka, Akito Oizumi, Masahiro
  Fukushima, Shoichiro Okita, Go~Chiba, Satoshi Sato, Masayuki Ohta, and Saerom
  Kwon.
\newblock Japanese evaluated nuclear data library version 5: {JENDL-5}.
\newblock {\em Journal of Nuclear Science and Technology}, 60(1):1--60, 2023.
\newblock \href {https://doi.org/10.1080/00223131.2022.2141903}
  {\path{doi:10.1080/00223131.2022.2141903}}.

\bibitem{Spellerberg1998}
S.~Spellerberg, P.~Reimer, G.~Blessing, H.~H. Coenen, and S.~M. Qaim.
\newblock Production of $^{55}$co and $^{57}$co via proton induced reactions on
  highly enriched $^{58}$ni.
\newblock {\em Applied Radiation and Isotopes}, 49(12):1519--1522, 1998.
\newblock \href {https://doi.org/10.1016/S0969-8043(97)10119-1}
  {\path{doi:10.1016/S0969-8043(97)10119-1}}.

\bibitem{Miyahara1998}
Hiroshi Miyahara, Atsushi Yoshida, Naoomi Ishikawa, and Chizuo Mori.
\newblock Simple source preparation and disintegration rate measurement of
  $^{56}$co and its use for efficiency calibration.
\newblock {\em Applied Radiation and Isotopes}, 49(9--11):1159--1164, 1998.
\newblock \href {https://doi.org/10.1016/S0969-8043(97)10038-0}
  {\path{doi:10.1016/S0969-8043(97)10038-0}}.

\bibitem{Khandaker2011}
Mayeen~UUddin Khandaker, Kwangsoo Kim, Manwoo Lee, Kyung~Sook Kim, and Guinyun
  Kim.
\newblock Excitation functions of (p,x) reactions on natural nickel up to 40
  mev.
\newblock {\em Nuclear Instruments and Methods in Physics Research Section B:
  Beam Interactions with Materials and Atoms}, 269:1140--1146, 2011.
\newblock \href {https://doi.org/10.1016/j.nimb.2011.02.082}
  {\path{doi:10.1016/j.nimb.2011.02.082}}.

\bibitem{Deiev2022NiPhotonuclear}
O.~S. Deiev, I.~S. Timchenko, S.~N. Olejnik, S.~M. Potin, V.~A. Kushnir, V.~V.
  Mytrochenko, S.~A. Perezhogin, and V.~A. Bocharov.
\newblock Photonuclear reactions natni$(\gamma,xn)^{57}$ni and
  natni$(\gamma,xn)^{56}$ni in the energy range $e_{\gamma}^{\mathrm{max}} =
  35$--$94$~mev.
\newblock {\em Nuclear Physics A}, 1028:122542, 2022.
\newblock \href {https://doi.org/10.1016/j.nuclphysa.2022.122542}
  {\path{doi:10.1016/j.nuclphysa.2022.122542}}.

\bibitem{Zaman2018NatNiPhotonuclear}
M.~Zaman, G.~Kim, H.~Naik, K.~Kim, M.~Shahid, M.~Nadeem, S.-G. Shin, and M.-H.
  Cho.
\newblock Flux-weighted average cross sections of natni$(\gamma,x)$ reactions
  with bremsstrahlung end-point energies of 55, 59, 61, and 65~mev.
\newblock {\em Nuclear Physics A}, 978:173--183, 2018.
\newblock \href {https://doi.org/10.1016/j.nuclphysa.2018.07.017}
  {\path{doi:10.1016/j.nuclphysa.2018.07.017}}.

\bibitem{Naik2020NatNiCo58}
H.~Naik, G.~Kim, T.~H. Nguyen, K.~Kim, S.-G. Shin, Y.~Kye, and M.-H. Cho.
\newblock Measurement of natni$(\gamma,xn)^{56,57}$ni and
  natni$(\gamma,pxn)^{58-55}$co reaction cross sections with bremsstrahlung
  end-point energies of 65 and 75~mev.
\newblock {\em Journal of Radioanalytical and Nuclear Chemistry}, 324:837--846,
  2020.
\newblock \href {https://doi.org/10.1007/s10967-020-07105-9}
  {\path{doi:10.1007/s10967-020-07105-9}}.

\bibitem{Timchenko2025Co58Photonuclear}
I.~S. Timchenko, O.~S. Deiev, S.~M. Olejnik, S.~M. Potin, V.~A. Kushnir, V.~V.
  Mytrochenko, S.~A. Perezhogin, and A.~Herz{\'a}{\v n}.
\newblock Cross section of the natni$(\gamma,pxn)^{58}$co reaction at the
  bremsstrahlung end-point energy of 35--94~mev.
\newblock {\em The European Physical Journal A}, 61:199, 2025.
\newblock \href {https://doi.org/10.1140/epja/s10050-025-01666-7}
  {\path{doi:10.1140/epja/s10050-025-01666-7}}.

\bibitem{Varlamov2010SnPhotonuclear}
V.~V. Varlamov, B.~S. Ishkhanov, V.~N. Orlin, and V.~A. Chetvertkova.
\newblock Evaluated cross sections of the $\sigma(\gamma, nx)$ and
  $\sigma(\gamma, 2nx)$ reactions on $^{112,114,116,117,118,119,120,122,124}$sn
  isotopes.
\newblock {\em Bulletin of the Russian Academy of Sciences: Physics},
  74(6):833--841, 2010.
\newblock \href {https://doi.org/10.3103/S1062873810060225}
  {\path{doi:10.3103/S1062873810060225}}.

\bibitem{Fultz1969}
S.~C. Fultz, B.~L. Berman, J.~T. Caldwell, R.~L. Bramblett, and M.~A. Kelly.
\newblock Photoneutron cross sections for $^{116,117,118,119,120,124}$sn and
  indium.
\newblock {\em Physical Review}, 186(4):1255--1270, 1969.
\newblock \href {https://doi.org/10.1103/PhysRev.186.1255}
  {\path{doi:10.1103/PhysRev.186.1255}}.

\bibitem{Utsunomiya2011}
H.~Utsunomiya, S.~Goriely, M.~Kamata, H.~Akimune, T.~Kondo, O.~Itoh,
  C.~Iwamoto, T.~Yamagata, H.~Toyokawa, Y.-W. Lui, H.~Harada, F.~Kitatani,
  S.~Goko, S.~Hilaire, and A.~J. Koning.
\newblock Photoneutron cross sections for $^{118}$--$^{124}$sn and the
  $\gamma$-ray strength function method.
\newblock {\em Physical Review C}, 84(5):055805, 2011.
\newblock \href {https://doi.org/10.1103/PhysRevC.84.055805}
  {\path{doi:10.1103/PhysRevC.84.055805}}.

\bibitem{MCNP63}
Joel~A. Kulesza, Terry~R. Adams, Jerawan~Chudoung Armstrong, Simon~R. Bolding,
  Forrest~B. Brown, Jeffrey~S. Bull, Timothy~Patrick Burke, Alexander~Rich
  Clark, Robert Arthur~III Forster, Jesse~Frank Giron, Tristan~Sumner Grieve,
  Colin~James Josey, Roger~L. Martz, Gregg~W. McKinney, Eric~John Pearson,
  Michael~Evan Rising, Clell~Jeffrey Solomon~Jr., Sriram Swaminarayan,
  Travis~John Trahan, Stephen~Christian Wilson, and Anthony~J. Zukaitis.
\newblock {MCNP}\textsuperscript{\textregistered} code version 6.3.0 theory \&
  user manual.
\newblock Technical Report LA-UR-22-30006, Los Alamos National Laboratory,
  September 2022.
\newblock OSTI ID: 1889957.
\newblock URL: \url{https://www.osti.gov/biblio/1889957}, \href
  {https://doi.org/10.2172/1889957} {\path{doi:10.2172/1889957}}.

\bibitem{SilverFirAttila2024}
{Silver Fir Software}.
\newblock {Attila}/{Attila4MC}: Deterministic transport and mcnp coupling
  software, version 10.3.
\newblock Computer software, 2024.
\newblock Silver Fir Software.

\bibitem{NISTSTAR2005}
M.~J. Berger, J.~S. Coursey, M.~A. Zucker, and J.~Chang.
\newblock {ESTAR}, {PSTAR}, and {ASTAR}: Computer programs for calculating
  stopping-power and range tables for electrons, protons, and helium ions
  (version 1.2.3).
\newblock [Online], 2005.
\newblock Available: \url{http://physics.nist.gov/Star}. Accessed: 2026-02-16.
  National Institute of Standards and Technology, Gaithersburg, MD.
\newblock URL: \url{http://physics.nist.gov/Star}.

\bibitem{Nusair2026}
O.~Nusair, D.~DeVries, and M.~Toro-Gonzalez.
\newblock Half-life measurements of 110sn, 113sn, 117msn, and 123msn produced
  via photon activation of natural tin.
\newblock {\em submitted to Nuclear Physics, Section A}, 2026.

\bibitem{BrunRademakers1997}
Rene Brun and Fons Rademakers.
\newblock {ROOT} -- an object-oriented data analysis framework.
\newblock {\em Nuclear Instruments and Methods in Physics Research Section A},
  389:81--86, 1997.
\newblock \href {https://doi.org/10.1016/S0168-9002(97)00048-X}
  {\path{doi:10.1016/S0168-9002(97)00048-X}}.

\bibitem{Timchenko2025Co5765Photonuclear}
I.~S. Timchenko, O.~S. Deiev, S.~M. Olejnik, S.~M. Potin, V.~A. Kushnir, V.~V.
  Mytrochenko, S.~A. Perezhogin, and A.~Herz{\'a}{\v n}.
\newblock Photoproduction of the $^{55\text{--}57}$co nuclei on natni at the
  bremsstrahlung end-point energy of 35--94~mev.
\newblock {\em Atomic Data and Nuclear Data Tables}, 160:101674, 2024.
\newblock \href {https://doi.org/10.1016/j.adt.2024.101674}
  {\path{doi:10.1016/j.adt.2024.101674}}.

\bibitem{BetheHeitler1934Stopping}
H.~Bethe and W.~Heitler.
\newblock On the stopping of fast particles and on the creation of positive
  electrons.
\newblock {\em Proceedings of the Royal Society of London. Series A},
  146(856):83--112, 1934.
\newblock \href {https://doi.org/10.1098/rspa.1934.0140}
  {\path{doi:10.1098/rspa.1934.0140}}.

\bibitem{Katz1951}
L.~Katz and A.~G.~W. Cameron.
\newblock The solution of x-ray activation curves for photonuclear cross
  sections.
\newblock {\em Canadian Journal of Physics}, 29:518--528, 1951.
\newblock \href {https://doi.org/10.1139/p51-056} {\path{doi:10.1139/p51-056}}.

\bibitem{Goryachev1970}
B.~I. Goryachev, B.~S. Ishkhanov, I.~M. Kapitonov, I.~M. Piskarev, V.~G.
  Shevchenko, and O.~A. Shevchenko.
\newblock {\em Soviet Journal of Nuclear Physics}, 11:141, 1970.

\bibitem{Fultz1974NiPhotonuclear}
S.~C. Fultz, R.~A. Alvarez, B.~L. Berman, and P.~Meyer.
\newblock Photonuclear reactions in nickel.
\newblock {\em Physical Review C}, 10:608--615, 1974.
\newblock \href {https://doi.org/10.1103/PhysRevC.10.608}
  {\path{doi:10.1103/PhysRevC.10.608}}.

\bibitem{Borodina2000}
S.~S. Borodina, A.~V. Varlamov, V.~V. Varlamov, B.~S. Ishkhanov, V.~I. Mokeev,
  and S.~I. Pavlov.
\newblock $^{54,56}$fe and $^{58,60}$ni $(\gamma,n)$, $(\gamma,p)$,
  $(\gamma,np)$, and $(\gamma,2n)$ reaction cross sections evaluation using the
  model of the gdr state decay channel competition phenomenological
  description.
\newblock Technical Report MSU-INP-2000-6/610, Skobeltsyn Institute of Nuclear
  Physics, Lomonosov Moscow State University, 2000.
\newblock URL: \url{http://www.sinp.msu.ru/en/preprint/7907}.

\bibitem{Ghosh2018}
Reetuparna Ghosh, Bioletty Lawriniang, Sylvia Badwar, Santhi~Sheela
  Yerraguntla, Haladhara Naik, Bhushankumar~J. Patil, Yeshwant Naik,
  Saraswatula~Venkata Suryanarayana, Betylda Jyrwa, and Srinivasan Ganesan.
\newblock Measurement and uncertainty propagation of the $(\gamma,n)$ reaction
  cross-section of $^{58}$ni and $^{59}$co at 15~mev bremsstrahlung.
\newblock {\em Radiochimica Acta}, 106(5):345--354, 2018.
\newblock \href {https://doi.org/10.1515/ract-2017-2855}
  {\path{doi:10.1515/ract-2017-2855}}.

\end{thebibliography}

\end{document}